\documentclass[twocolumn,prb,showpacs,preprintnumbers,amsmath,amssymb,floatfix,superscriptaddress]{revtex4}
\usepackage{graphicx}
\usepackage{amsmath}
\usepackage{amssymb}
\usepackage{color}

\begin{document}

\title{Solution of the tunneling-percolation problem in the nanocomposite regime}

\author{G. Ambrosetti}\email{gianluca.ambrosetti@a3.epfl.ch}\affiliation{LPM, Ecole Polytechnique F\'ed\'erale de Lausanne, Station 17, CH-1015 Lausanne, Switzerland}\affiliation{ICIMSI, University of Applied Sciences of
Southern Switzerland, CH-6928 Manno, Switzerland}
\author{C. Grimaldi}\email{claudio.grimaldi@epfl.ch}\affiliation{LPM, Ecole Polytechnique F\'ed\'erale de
Lausanne, Station 17, CH-1015 Lausanne, Switzerland}
\author{I. Balberg}\affiliation{The Racah Institute of Physics, The Hebrew University, Jerusalem 91904, Israel}
\author{T. Maeder}\affiliation{LPM, Ecole Polytechnique F\'ed\'erale de
Lausanne, Station 17, CH-1015 Lausanne, Switzerland}
\author{A. Danani}\affiliation{ICIMSI, University of Applied Sciences of
Southern Switzerland, CH-6928 Manno, Switzerland}
\author{P. Ryser}\affiliation{LPM, Ecole Polytechnique F\'ed\'erale de
Lausanne, Station 17, CH-1015 Lausanne, Switzerland}



\begin{abstract}
We noted that the tunneling-percolation framework is quite well understood at the extreme cases of percolation-like and hopping-like behaviors but that the intermediate regime has not been previously discussed, in spite of its relevance to the intensively studied electrical properties of nanocomposites. Following that we study here the conductivity of
dispersions of particle fillers inside an insulating matrix by taking into account explicitly the
filler particle shapes and the inter-particle electron tunneling process. We show that the main features of
the filler dependencies of the nanocomposite conductivity can be reproduced without introducing any \textit{a priori}
imposed cut-off in the inter-particle conductances, as usually done in the percolation-like interpretation
of these systems. Furthermore, we demonstrate that our numerical results are fully reproduced by the critical path
method, which is generalized here in order to include the particle filler shapes. By exploiting this method, we
provide simple analytical formulas for the composite conductivity valid for many regimes of interest.
The validity of our formulation is assessed by reinterpreting existing experimental results on nanotube, nanofiber, nanosheet and nanosphere composites and by extracting the
characteristic tunneling decay length, which is found to be within the expected range of its values.
These results are concluded then to be not only useful for the understanding of the intermediate regime but
also for tailoring the electrical properties of nanocomposites.
\end{abstract}
\maketitle

\section{Introduction}
\label{intro}

The inclusion of nanometric conductive fillers such as carbon nanotubes \cite{Bauhofer2009}, nanofibers \cite{Al-Saleh2009}, and graphene
\cite{Eda2009,Stankovich2006} into insulating matrices allows to obtain electrically conductive
nanocomposites with unique properties which are widely investigated and have several technological
applications ranging from antistatic coatings to printable electronics \cite{Sekitani2009}.
A central challenge in this domain is to create composites with an
overall conductivity $\sigma$ that can be controlled by the volume fraction $\phi$, the shape of the
conducting fillers, their dispersion in the insulating matrix, and the local inter-particle electrical connectedness.
Understanding how these local properties affect the composite conductivity is therefore the ultimate goal of
any theoretical investigation of such composites.

A common feature of most random insulator-conductor mixtures is the sharp increase of $\sigma$ once a critical
volume fraction $\phi_c$ of the conductive phase is reached. This transition is generally interpreted in the
framework of percolation theory \cite{Kirkpatrick1973,Stauffer1994,Sahimi2003} and associated with the formation
of a cluster of electrically connected filler particles that spans the entire sample.
The further increase of $\sigma$ for $\phi>\phi_c$ is likewise understood
as the growing of such a cluster.  In the vicinity of $\phi_c$, this picture implies a power-law behavior of the
conductivity of the form
\begin{equation}
\label{plaw}
\sigma\propto (\phi-\phi_c)^t,
\end{equation}
where $t$ is a critical exponent.
Values of $t$ extracted from experiments range from its expected universal value for three-dimensional percolating systems, $t\simeq 2$, up
to $t\simeq 10$, with little or no correlation to the critical volume fraction $\phi_c$,\cite{Vionnet2005} or the shape of the conducting fillers \cite{Bauhofer2009}.

In the dielectric regime of a system of nanometric conducting particles embedded in a continuous
insulating matrix, as is the case for conductor-polymer
nano-composites,\cite{Sheng1978,Balberg1987b,Paschen1995,Li2007} the particles do not physically touch
each other, and the electrical connectedness is established through tunneling
between the conducting filler particles. In this situation, the basic assumptions of
percolation theory are, a priori, at odds with the inter-particle
tunneling mechanism.\cite{Balberg2009}
Indeed, while percolation requires the introduction of some sharp cut-off in the inter-particle
conductances, i.e., the particles are either connected (with given non-zero inter-particle conductances)
or disconnected,\cite{Stauffer1994,Sahimi2003} the tunneling between particles is a continuous
function of inter-particle distances.
Hence, the resulting tunneling conductance, which decays exponentially with these distances, does not
imply any sharp cut-off or threshold.

Quite surprisingly, this fundamental incompatibility has hardly been discussed in the
literature,\cite{Balberg2009} and basically all the measured conductivity dependencies on the
fractional volume content of the conducting phase, $\sigma(\phi)$, have been interpreted in terms
of Eq.~\eqref{plaw} assuming the ``classical'' percolation  behavior.\cite{Stauffer1994,Sahimi2003}
In this article, we show instead that the inter-particle tunneling explains well all the main features
of $\sigma(\phi)$ of nanocomposites without imposing any {\it a priori} cut-off, and that it
provides a much superior description of $\sigma(\phi)$ than the ``classical'' percolation formula
\eqref{plaw}.

In order to specify our line of reasoning and
to better appreciate the above mentioned incompatibility, it is instructive to consider a system of particle
dispersed in an insulating continuum with a tunneling conductance between two of them, $i$ and $j$, given by:
\begin{equation}
\label{eq:tunneling}
g_{ij}=g_0\exp\!\left(-\frac{2\delta_{ij}}{\xi}\right),
\end{equation}
where $g_0$ is a constant, $\xi$ is the characteristic tunneling length, and $\delta_{ij}$ is the
minimal distance between the two particle surfaces. For spheres of diameter $D$, $\delta_{ij}=r_{ij}-D$
where $r_{ij}$ is the center-to-center distance. There are two extreme cases for which the
resulting composite conductivity has qualitatively different behaviors which can be easily described.
In the first case the particles are so large that $\xi/D\rightarrow 0$. It becomes then clear
from Eq.\eqref{eq:tunneling} that the conductance between two particles is non-zero
only when they essentially touch each other.
Hence, removing particles from the random closed packed limit is equivalent to remove tunneling
bonds from the system, in analogy to sites removal in a site percolation problem in the lattice.\cite{Stauffer1994,Sahimi2003}
The system conductivity will have then a percolation-like behavior as in Eq.~\eqref{plaw}
with $t\simeq 2$ and $\phi_c$ being the corresponding percolation threshold.\cite{Johner2008}
The other extreme case is that of sites ($D/\xi\rightarrow 0$) randomly dispersed in the continuum.
In this situation, a variation of the site density $\rho$ does not change the connectivity between the particles and
its only role is to vary the distances $\delta_{ij}=r_{ij}$ between the sites.\cite{Balberg2009,Shklovskii1984}
The corresponding $\sigma$ behavior was solved by using the critical path (CP) method\cite{Ambegaokar1971}
in the context of hopping in amorphous semiconductors yielding
$\sigma\propto \exp[-1.75/(\xi\rho^{1/3})]$.\cite{Seager1974,Overhof1989}
For sufficiently dilute system of impenetrable spheres this relation can be
generalized to $\sigma\propto \exp[-1.41D/(\xi\phi^{1/3})]$.\cite{Balberg2009}
It is obvious then from the above discussion that the second case is the low density limit of
the first one, but it turns out that the variation of $\sigma(\phi)$ between the two types of situations,
which is definitely pertinent to nanocomposites, has not been studied thus far.

Following the above considerations we turned to study here the $\sigma(\phi)$ dependencies
by extending the low-density (hopping-like)
approach to higher densities than those used previously.\cite{Shklovskii1984,Seager1974,Overhof1989}
Specifically, we shall present numerical results obtained by using the global tunneling network (GTN) model,
where the conducting fillers form a network of globally connected sites via tunneling processes.
This model has already been introduced in Ref.[\onlinecite{Ambrosetti2009}] for the case of impenetrable spheres,
but here we shall generalize it in order to describe also anisotropic fillers such as rod-like and
plate-like particles, as to apply to cases of recent interest ({\it i.e.}, nanotube, nanofiber, nanosheet, and
graphene composites). In particular, the large amount of published experimental data on these systems
allows us to test the theory and to extract the values of microscopic parameters directly from
macroscopic data on the electrical conductivity.

The structure of the paper is as follows. In Sec.~\ref{gen} we describe how we generate particle
dispersions and in Sec.~\ref{cond} we calculate numerically the composite conductivities within the
GTN model and compare them with the conductivities obtained by the CP approximation. In Sec.~\ref{critical}
we present our results on the critical tunneling distance which are used in Sec.~\ref{formulas} to obtain
analytical formulas for the composite conductivity. These are applied in Sec.~\ref{compa} to several
published data on nanocomposites to extract the tunneling distance. Section \ref{concl} is devoted to
discussions and conclusions.

\section{sample generation}
\label{gen}

In modeling the conductor-insulator composite morphology, we treat the conducting fillers as identical
impenetrable objects dispersed in a continuous insulating medium, with no interactions between the
conducting and insulating phases. As pointed out above, in order to relate to systems of recent
interest we describe filler particle shapes that vary from rod-like (nanotubes) to plate-like
(graphene). This is done by employing impenetrable spheroids (ellipsoids of revolution) ranging from the extreme
prolate ($a/b\gg 1$) to the extreme oblate limit ($a/b\ll 1$),
where $a$ and $b$ are the spheroid polar and equatorial semi-axes respectively.

\begin{figure*}[t]
    \begin{center}
\includegraphics[scale=0.7, clip='true']{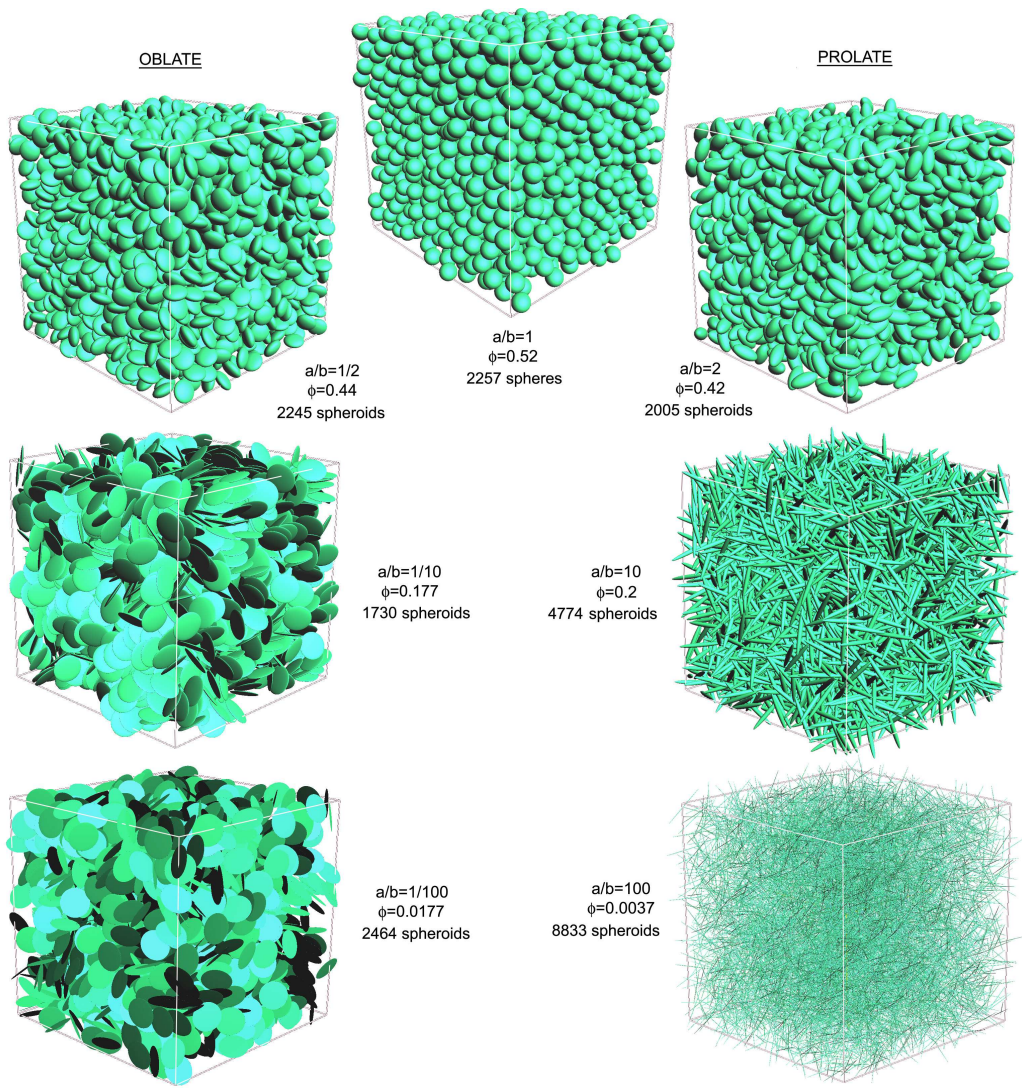}
  \caption{(Color online) Examples of distributions of impenetrable spheres and
  spheroids of different aspect-ratios $a/b$ and volume fraction $\phi$ generated by the algorithms used
  in the present work.}\label{Fig1}
  \end{center}
\end{figure*}

We generate dispersions of non-overlapping spheroids by using an extended version of a previously described
algorithm\cite{Ambrosetti2008} which allows to add spheroids into a cubic cell with periodic boundary conditions
through random sequential addition (RSA).\cite{Sherwood1997} Since the configurations obtained via RSA are
non-equilibrium ones,\cite{Torquato2001,Miller2009} the RSA dispersions were relaxed via Monte Carlo (MC) runs,
where for each spheroid a random displacement of its center and a random rotation of its axes \cite{Frenkel1985}
were attempted, being accepted only if they did not give rise to an overlap with any of its neighbors.
Equilibrium was considered attained when the ratio between the number of accepted trial moves versus the number
of rejected ones had stabilized. Furthermore, to obtain densities beyond the ones obtainable with RSA, a high
density generation procedure \cite{Ambrosetti2009,Miller1990} was implemented where in combination with MC
displacements the particles were also inflated.
The isotropy of the distributions was monitored by using the nematic order parameter as described in
Ref.[\onlinecite{Schilling2007}].

Figure \ref{Fig1} shows examples of the so-generated distributions for spheroids with different
aspect-ratios $a/b$ and volume fractions $\phi=V\rho$, where $V=4\pi a b^2/3$ is the volume of a single spheroid and $\rho$ is the particle number density.

\section{determination of the composite conductivity by the GTN and CP methods}
\label{cond}

In considering the overall conductivity arising in such composites, we attributed to each spheroid pair
the tunneling conductance given in Eq.~\eqref{eq:tunneling} where, now, for $a/b\neq 1$ the inter-particle
distance $\delta_{ij}$ depends also on the relative orientation of the spheroids. The $\delta_{ij}$
values were obtained here from the numerical procedure described in Ref.[\onlinecite{Ambrosetti2008}].
On the other hand, in writing Eq.~\eqref{eq:tunneling} we neglect any energy difference between
spheroidal particles and disregard
activation energies since, in general, these contributions can be ignored at relatively high
temperatures,\cite{Shklovskii1984,Sheng1983} which is the case of interest here.
For the specific case of extreme prolate objects ($a/b\gg 1$) the regime
of validity of this approximation has been studied in Ref.[\onlinecite{Hu2006}].

The full set of bond conductances given by Eq.~\eqref{eq:tunneling} was mapped as a resistor network
with $g_0=1$ and the overall conductivity was calculated through numerical decimation of the resistor network.\cite{Johner2008,Fogelholm1980}
To reduce computational times of the decimation procedure to
manageable limits, an artificial maximum distance was
introduced in order to reject negligibly small bond conductances. It
is important to note that this artifice is not in conflict with the
rationale of the GTN model, since the cutoff it implies neglects
conductances which are completely irrelevant for the global system
conductivity. We chose the maximum distance to be generally fixed and
equal to four times the spheroid major axis (i.e. $a$ in the prolate
case and $b$ in the oblate case), which is equivalent to reject
inter-particle conductances below $e^{-60}$ for $\xi/D=1/15$ case (and
considerably less for smaller $\xi$ values). However, for the high aspect-ratios
and high densities the distance had to be reduced.
Moreover, since the maximum distance implies in turn an artificial
geometrical percolation threshold of the system, for the high
aspect-ratios, at low volume fractions the distance had to be
increased to avoid this effect. By comparing the results with the ones
obtained with significantly larger maximum distances we verified that the
effect is undetectable.

\begin{figure*}[t]
    \begin{center}
\includegraphics[scale=0.6, clip='true']{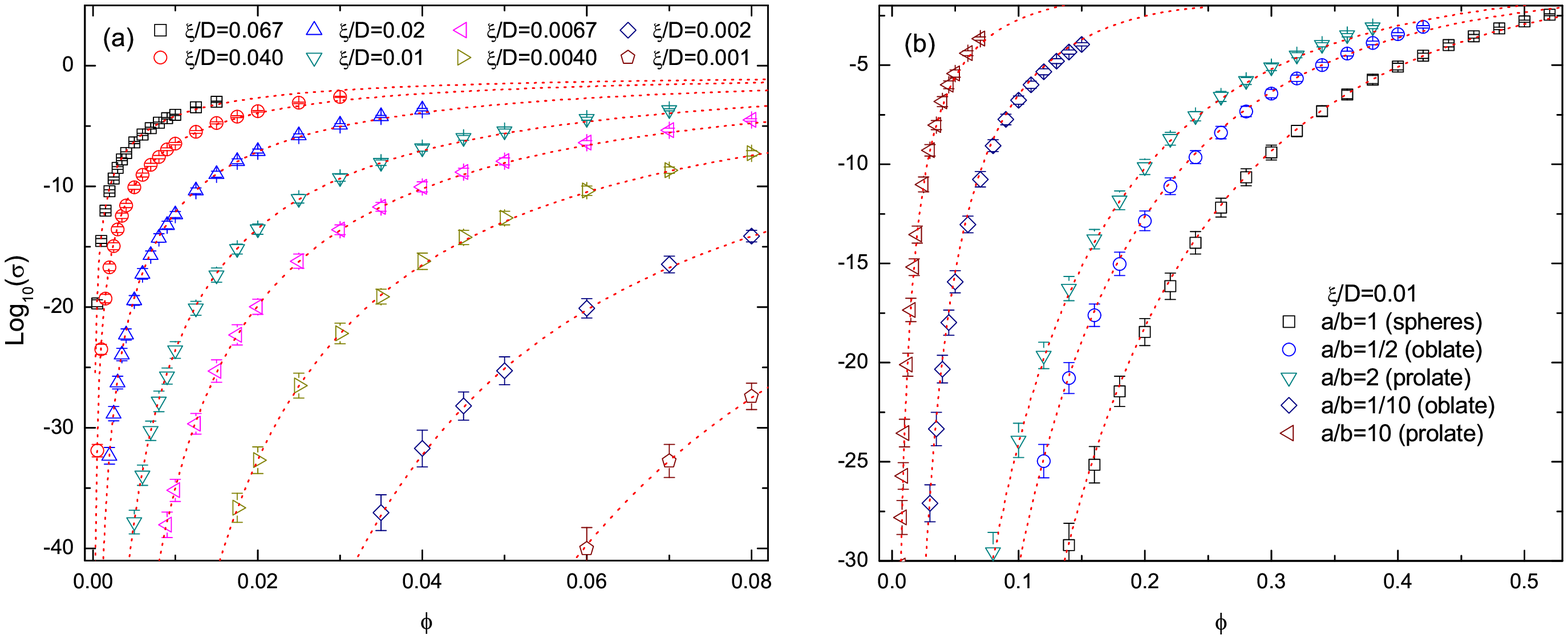}
  \caption{(Color online) The results of our GTN and CP calculations.
(a): Volume fraction $\phi$ dependence of the tunneling conductivity $\sigma$ for a system
of aspect-ratio $a/b=10$ hard prolate spheroids with different characteristic tunneling distances $\xi/D$ with $D=2a$.
Results from Eq.~\eqref{eq:sigmaSP} (with $\sigma_0=0.179$) are displayed by dotted lines.
(b): Tunneling conductivity in a system of hard spheroids with different aspect-ratios $a/b$ and $\xi/D=0.01$,
with $D=2\max(a,b)$. Dotted lines: results from Eq.~\eqref{eq:sigmaSP} with $\sigma_{0}=0.124$ for $a/b=2$,
$\sigma_{0}=0.099$ for $a/b=1/2$, $\sigma_{0}=0.351$ for $a/b=1/10$, and $\sigma_{0}=0.115$ for $a/b=1$.}\label{Fig2}
  \end{center}
\end{figure*}

In Fig.~\ref{Fig2}(a) we show the so-obtained conductivity $\sigma$ values (symbols) as
a function of the volume fraction $\phi$ of prolate spheroids with aspect-ratio $a/b=10$ and
different values of $\xi/D$, where $D=2\max(a,b)$.
Each symbol is the outcome of $N_R=200$ realizations of a system of $N_P\sim1000$ spheroids.
The logarithm average of the results was considered since, due to the exponential dependence of Eq.~\eqref{eq:tunneling}, the distribution of the computed conductivities was approximately of
the log-normal form.\cite{Strelniker2005} The strong reduction of $\sigma$ for decreasing
$\phi$ shown in Fig.~\ref{Fig2}(a) is a direct consequence of the fact that as $\phi$ is reduced,
the inter-particle distances get larger, leading in turn to a reduction of the local tunneling
conductances [Eq.~\eqref{eq:tunneling}].
In fact, as shown in Fig.~\ref{Fig2}(b), this reduction depends strongly on
the shape of the conducting fillers. Specifically, as the shape anisotropy of the particles is enhanced, the composite
conductivity drops for much lower values of $\phi$ for a fixed $\xi$.

\begin{figure}[t]
    \begin{center}
\includegraphics[scale=0.3, clip='true']{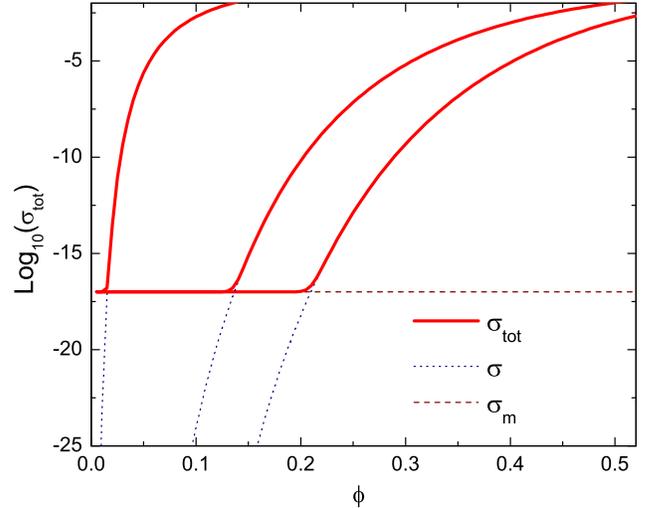}
  \caption{(Color online) Schematic illustration of the tunneling conductivity crossover for the cases
  $a/b=1$, $a/b=2$, and $a/b=10$.}\label{Fig3}
  \end{center}
\end{figure}

Having the above result we turn now to show that the strong dependence of $\sigma(\phi)$ on $a/b$
and $\xi$ in Fig.~\ref{Fig2}
can be reproduced by CP method \cite{Shklovskii1984,Ambegaokar1971,Seager1974,Overhof1989} when
applied to our system of impenetrable spheroids. For the tunneling conductances of Eq.~\eqref{eq:tunneling},
this method amounts to keep only the subset of conductances $g_{ij}$ having $\delta_{ij}\leq\delta_c$,
where $\delta_c$, which defines the characteristic conductance $g_c=g_0\exp(-2\delta_c/\xi)$,
is the largest among the $\delta_{ij}$ distances, such that the so-defined subnetwork
forms a conducting cluster that span the sample. Next, by assigning $g_c$ to all the (larger) conductances
of the subnetwork, a CP approximation for $\sigma$ is
\begin{equation}
\label{eq:sigmaSP}
\sigma\simeq\sigma_0\exp\left[-\frac{2\delta_c(\phi,a,b)}{\xi}\right],
\end{equation}
where $\sigma_0$ is a pre-factor proportional to $g_0$.
The significance of Eq.~\eqref{eq:sigmaSP} is that it reduces the conductivity of a distribution of
hard objects that are electrically connected by tunneling to the computation of
the geometrical ``critical" distance $\delta_c$.
In practice, $\delta_c$ can be obtained by coating each impenetrable spheroid  with a
penetrable shell of constant thickness $\delta/2$, and by considering two spheroids as connected if
their shells overlap. $\delta_c$ is then the minimum value of $\delta$ such that, for a given $\phi$,
a cluster of connected spheroids spans the sample.

To extract $\delta_c$ we follow the route outlined in Ref.~[\onlinecite{Ambrosetti2008}] with the extended
distribution generation algorithm described in Sec.~\ref{gen}. Specifically, we calculated the spanning
probability as a function of $\phi$ for fixed $a/b$ and $\delta_c$ by recording the frequency of appearance
of a percolating cluster over a given number of realizations $N_R$.
The realization number varied from $N_R=40$ for the smallest values of $\delta_c$ up to $N_R=500$ for
the largest ones. Each realization involved distributions of $N_P\sim 2000$ spheroids, while for high
aspect-ratio prolate spheroids this number increased to $N_P\sim 8000$ in order to be able to maintain
the periodic boundary conditions on the simulation cell. Relative errors on $\delta_c$
were in the range of a few per thousand.

Results of the CP approximation are reported in Fig.~\ref{Fig2} by dotted lines. The agreement with
the full numerical decimation of the resistor network is excellent for all values of $a/b$ and $\xi/D$
considered. This observation is quite important since it shows that the CP method is valid also beyond
the low-density regime, for which the
conducting fillers are effectively point particles, and that it can be successfully used for systems of particles
with impenetrable volumes.  Besides the clear practical advantage of evaluating $\sigma$ via the geometrical
quantity $\delta_c$ instead of solving the whole resistor network, the CP
approximation is found then, as we shall see in the next section, to allow the full understanding of the filler
dependencies of $\sigma$ and to identify asymptotic formulas for many regimes of interest.

Before turning to the analysis of the next section, it is important at this point to discuss the following issue.
As shown in Fig.~\ref{Fig2}, the GTN scenario predicts, in principle, an
indefinite drop of $\sigma$ as $\phi\rightarrow 0$ because, by construction, there is not an imposed cut-off in the
inter-particle conductances. However, in real composites, either the lowest measurable conductivity is
limited by the experimental set-up,\cite{Balberg2009} or it is given by the intrinsic conductivity $\sigma_{\rm m}$
of the insulating matrix, which prevents an indefinite drop of $\sigma$.
For example, in polymer-based composites $\sigma_{\rm m}$ falls typically in the range of
$\sigma_{\rm m}\simeq 10^{-13}\div 10^{-18}$~S/cm, and it originates from ionic impurities or displacement
currents.\cite{Blyte2005} Since the contributions from the polymer and the inter-particle tunneling
come from independent current paths, the total conductivity (given by the polymer and the inter-particle
tunneling) is then simply $\sigma_{\rm tot}=\sigma_{\rm m}+\sigma$.\cite{notecross} As illustrated in
Fig.~\ref{Fig3}, where $\sigma_{\rm tot}$ is plotted for $a/b=1$, $2$, and $10$ and for
$\sigma_{\rm m}/\sigma_0=10^{-17}$, the $\phi$-dependence of $\sigma_{\rm tot}$ is characterized
by a cross-over concentration $\phi_c$ below which $\sigma_{\rm tot}\simeq \sigma_{\rm m}$.
As seen in this figure, fillers with larger shape-anisotropy entail lower values of $\phi_c$, consistently
with what is commonly observed.\cite{Bauhofer2009,Ota1997,Nagata1999,Lu2006}
We have therefore that the main features of nanocomposites
(drop of $\sigma$ for decreasing $\phi$, enhancement of $\sigma$ at fixed $\phi$ for larger
particle anisotropy, and a characteristic $\phi_c$ below which the conductivity matches that of the insulating
phase) can be obtained without invoking any microscopic cut-off, leading therefore to a radical change
of perspective from the classical percolation picture. In particular, in the present context,
the conductor-insulator transition is no longer described as a true percolation transition
(characterized by a critical behavior of $\sigma$ in the vicinity of a definite percolation
threshold, \textit{i.e.}, Eq.\eqref{plaw}), but rather
as a cross-over between the inter-particle tunneling conductivity and the insulating matrix conductivity.

\section{CP determination of the critical distance $\delta_c$ for spheroids}
\label{critical}

The importance of the CP approximation for the understanding of the filler dependencies of $\sigma$
is underscored by the fact that, as discussed below, for sufficiently elongated prolate
and for sufficiently flat oblate spheroids,
as well as for spheres, simple relations exist that allow to estimate the value of $\delta_c$ with good accuracy.
In virtue of Eq.~\eqref{eq:sigmaSP} this means that we can formulate explicit relations between $\sigma$ and
the shapes and concentration of the conducting fillers.

\subsection{prolate spheroids}
\label{prolate}

\begin{figure*}[t]
    \begin{center}
  \includegraphics[scale=0.6, clip='true']{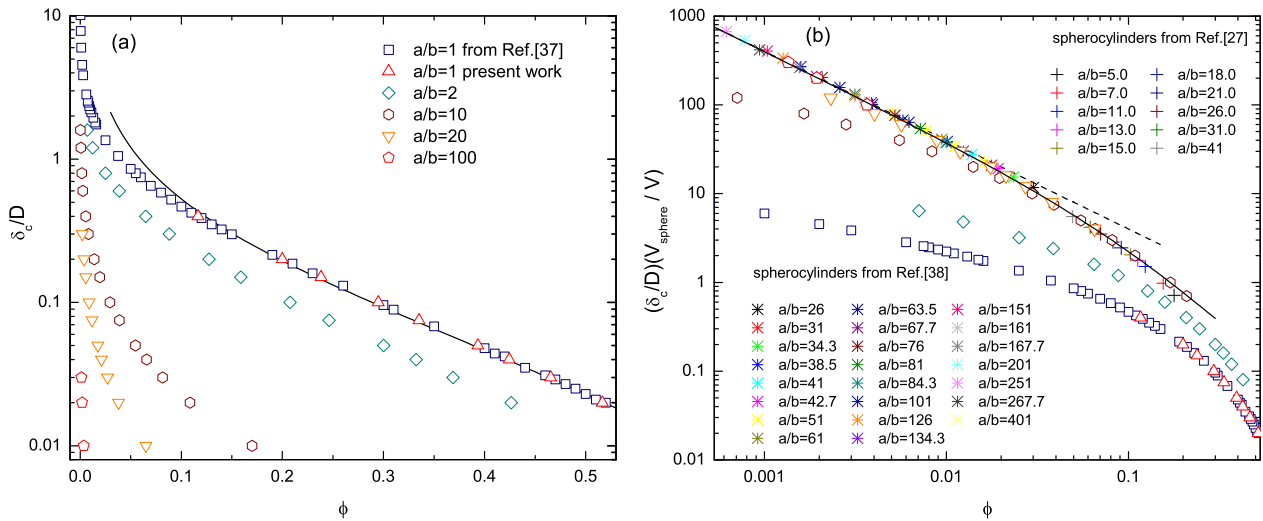}
\caption{(Color online) (a): the $\delta_c/D$ dependence on the volume fraction $\phi$
for impenetrable prolate spheroids with $a/b=1$, $2$, $10$, $20$, and $100$.
For $a/b=1$ our results are plotted together with those of Ref.~[\onlinecite{Heyes2006}].
The solid line is Eq.~\eqref{eq:deltasphere}.
(b): re-scaled critical distances versus $\phi$ for prolate spheroids as well as for the impenetrable
spherocylinders of Refs.~[\onlinecite{Schilling2007,Berhan2007}]. The dashed line follows Eq.~\eqref{eq:deltapro}
and the solid line follows Eq.~\eqref{eq:deltapro2} }\label{Fig4}
  \end{center}
\end{figure*}

Let us start with prolate ($a/b>1$) spheroids. In Fig.~\ref{Fig4}(a) we present the calculated values of
$\delta_c/D$ as a function of the volume fraction $\phi$ for spheres ($a/b=1$, together with the results of
Ref.~[\onlinecite{Heyes2006}]) and for
$a/b=2$, $10$, $20$, and $100$. In the log-log plot of Fig.~\ref{Fig4}(b) the same data are
displayed with $\delta_c/D$ multiplied by the ratio $V_{\rm sphere}/V=(a/b)^2$,
where $V_{\rm sphere}=\pi D^3/6$ is the volume of a sphere with diameter equal to the major axis of the
prolate spheroid and $V=4\pi a b^2/3$ is the volume of the spheroid itself.
For comparison, we also plot in Fig.~\ref{Fig4}(b) the results for impenetrable
spherocylinders of Refs.~[\onlinecite{Schilling2007,Berhan2007}]. These are formed by cylinders of radius $R$
and length $L$, capped by hemispheres of radius $R$, so that $a=R+L/2$ and $b=R$, and
$V_{\rm sphere}/V=(a/b)^{3}/[(3/2)(a/b)-2]\simeq (2/3)(a/b)^{2}$ for $a/b\gg 1$.
As it is apparent, for sufficiently large values of $a/b$ the simple re-scaling transformation
collapses both spheroids and spherocylinders data into a single curve. This holds true as long as the
aspect-ratio of the spheroid plus the penetrable shell $(a+\delta_c/2)/(b+\delta_c/2)$ is larger than
about $5$. In addition, for $\phi \lesssim 0.03$ the collapsed data are well approximated by
$\delta_c V_{\rm sphere}/V/D=0.4/\phi$ [dashed line in Fig.~\ref{Fig4}(b)], leading to the following
asymptotic formula:
\begin{equation}
\label{eq:deltapro}
\delta_c/D\simeq \frac{\gamma (b/a)^2}{\phi},
\end{equation}
where $\gamma=0.4$ for spheroids and $\gamma=0.6$ for spherocylinders. Equation \eqref{eq:deltapro}
is fully consistent with the scaling law of Ref.~[\onlinecite{Kyrylyuk2008}] that was obtained from the
second-virial approximation for semi-penetrable spherocylinders, and
it can be understood from simple excluded volume effects.
Indeed, in the asymptotic regime $a/b\gg 1$ and for $\delta_c/a\ll 1$, the filler density $\rho$
(such that a percolating cluster of connected semi-penetrable spheroids with penetrable shell $\delta_c$
is formed) is given by $\rho=1/\Delta V_{\rm exc}$.\cite{Berhan2007,Balberg1984,Bug1986}
Here, $\Delta V_{\rm exc}$ is the excluded volume of a randomly oriented semi-penetrable object minus
the excluded volume of the impenetrable object.
As shown in the Appendix, for both spheroids and
spherocylinder particles this becomes $\Delta V_{\rm exc}\simeq 2\pi a^2\delta_c$,
leading therefore to Eq.~\eqref{eq:deltapro} with $\gamma=1/3$ for spheroids and $\gamma=1/2$ for
spherocylinders.\cite{noteshoot}

It is interesting to notice that in Fig.~\ref{Fig4}(b) the re-scaled data for $\phi\gtrsim 0.03$
deviate from Eq.~\eqref{eq:deltapro} but still follow a common curve. We have found that this common trend
is well fitted by an empirical generalization of Eq.~\eqref{eq:deltapro}:
\begin{equation}
\label{eq:deltapro2}
\delta_c/D\simeq \frac{\gamma (b/a)^2}{\phi(1+8\phi)},
\end{equation}
which applies to all values of $\phi$ provided that $(a+\delta_c/2)/(b+\delta_c/2)\gtrsim 5$
[solid lines in Fig.~\ref{Fig4}(b)].

\subsection{oblate spheroids}
\label{oblate}

\begin{figure*}[t]
    \begin{center}
  \includegraphics[scale=0.3, clip='true']{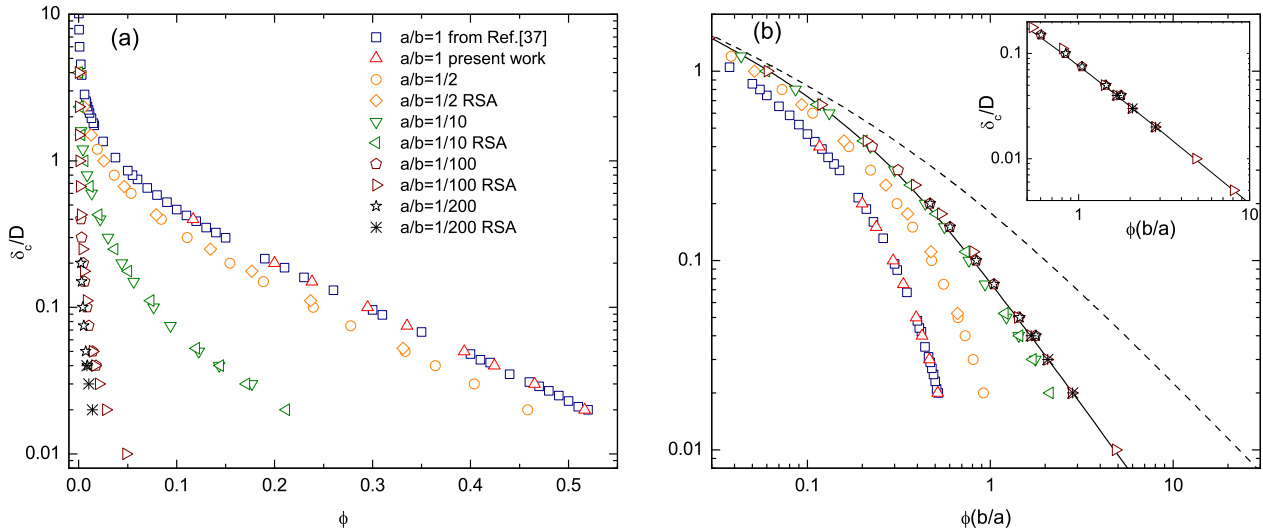}
  \caption{(Color online) (a): the $\delta_c/D$ dependence on the volume fraction $\phi$ for impenetrable
  oblate spheroids with $a/b=1$, $1/2$, $1/10$, $1/100$, and $1/200$.
  Results obtained by RSA alone are also presented.
(b): our $\delta_c/D$ values plotted versus the re-scale volume fraction
  $\phi (b/a)$. The dashed line follows Eq.~\eqref{eq:PcParallel} and the solid
  line follows Eq.~\eqref{eq:PcOblates}. Inset: the asymptotic behavior for $\delta_c/D<0.1$.}\label{Fig5}
  \end{center}
\end{figure*}

Let us now turn to the case of oblate spheroids. The numerical results for $\delta_c$ as a function of
the volume fraction $\phi$ are displayed in Fig.~\ref{Fig5}(a) for $a/b=1$, $1/2$, $1/10$, $1/100$,
and $1/200$. Now, as opposed to prolate fillers, almost all of the experimental results on nanocomposites,
such as graphene,\cite{Stankovich2006} that contain oblate filler with high shape anisotropy are at volume
fractions for which a corresponding hard spheroid fluid at equilibrium would already be in the nematic phase.
For oblate spheroids with $a/b=1/10$ the isotropic-nematic (I-N) transition is at
$\phi_{\rm I-N}\sim 0.185$,\cite{Allen1989} while for lower $a/b$ values the transition may be
estimated from the results on infinitely thin hard disks:\cite{Eppenga1984}  $\phi_{\rm I-N}\sim0.0193$ for
$a/b=1/100$ and $\phi_{\rm I-N}\sim0.0096$ for $a/b=1/200$.
However, in real nanocomposites the transition to the nematic phase is hampered by the viscosity of the
insulating matrix and these systems are inherently out of equilibrium.\cite{kharchenko}
In order to maintain global isotropy also for $\phi>\phi_{\rm I-N}$, we generated oblate
spheroid distributions with RSA alone.
The outcomes are again displayed in Fig.~\ref{Fig5} and one can appreciate that the difference with
the equilibrium results for $\phi<\phi_{\rm I-N}$ is quite small and negligible for the present aims.

In analogy to what we have done for the case of prolate objects, it would be useful to find a scaling
relation permitting to express the $\phi$-dependence of $\delta_c/D$ also for oblate spheroids, at least
for the $a/b\ll 1$ limit, which is the one of practical interest. To this end, it is instructive to consider
the case of perfectly parallel spheroids which can be easily obtained from general result for aligned
penetrable objects.\cite{Balberg1985} For infinitely thin parallel hard disks
of radius $b$ one therefore has $V_{\rm exc}^{\|}=2.8/\rho$, where $V_{\rm exc}^{\|}=(4/3)\pi b^{3}[12(\delta_c/D)+6\pi(\delta_c/D)^{2}+8(\delta_c/D)^{3}]$ is the
excluded volume of the plate plus the penetrable shell of critical thickness $\delta_c/2$.
Assuming that this holds true also for hard-core-penetrable-shell oblate spheroids with a sufficiently thin
hard core, we can then write
\begin{equation}
\label{eq:PcParallel}
12(\delta_c/D)+6\pi(\delta_c/D)^{2}+8(\delta_c/D)^{3}\simeq \frac{2.8}{\phi (b/a)},
\end{equation}
which implies that $\delta_c/D$ depends solely on $\phi (b/a)$. As shown in the Appendix, where
the excluded volume of an isotropic orientation of oblate spheroids is reported, also the
second-order virial approximation gives $\delta_c/D$ as a function of $\phi(a/b)$ for $a/b\ll 1$.
Hence, although Eq.~\eqref{eq:PcParallel} and Eq.~\eqref{phiobl} are not expected to be quantitatively
accurate, they suggest nevertheless a possible way of rescaling the data of Fig.~\ref{Fig5}(a).
Indeed, as shown in Fig.~\ref{Fig5}(b), for sufficiently high shape anisotropy the data of $\delta_c/D$
plotted as a function of $\phi(a/b)$ collapse into a single curve (the results for $a/b=1/100$ and $a/b=1/200$
are completely superposed). Compared to Eq.~\eqref{eq:PcParallel}, which behaves as
$\delta_c/D\propto [\phi(b/a)]^{-1}$ for $\delta_c/D\ll 1$ (dashed line), the re-scaled data
in the log-log plots of Fig.~\ref{Fig5}(b) still follow a straight line in the same range of $\delta_c/D$
values but with a slightly sharper slope, suggesting a power-law dependence on $\phi(a/b)$.
Empirically, Eq.~\eqref{eq:PcParallel} does indeed reproduce then the $a/b\ll 1$ asymptotic behavior
by simply modifying the small $\delta_c/D$ behavior as follows:
\begin{equation}
\label{eq:PcOblates}
12\alpha(\delta_c/D)^{\beta}+6\pi(\delta_c/D)^{2}+8(\delta_c/D)^{3} \simeq \frac{2.8}{\phi (b/a)},
\end{equation}
where $\alpha=1.54$ and $\beta=3/4$. When plotted against our data, Eq.~\eqref{eq:PcOblates}
(solid line) provides an accurate approximation for $\delta_c/D$ in the whole range of $\phi(a/b)$
for $a/b< 1/100$.  Moreover, by retaining the dominant contribution of Eq.~\eqref{eq:PcOblates}
for  $\delta_c/D<0.1$, we arrive at (inset of Fig.~\ref{Fig5}(b):
\begin{equation}
\label{eq:deltaOblates}
\delta_c/D\simeq\left[\frac{0.15(a/b)}{\phi}\right]^{4/3},
\end{equation}
which applies to all cases of practical interest for plate-like filler particles ($a/b\ll 1$ and
$\delta_c/D\ll 1$).

\subsection{spheres}
\label{spheres}

Let us conclude this section by providing an accurate expression for $\delta_c/D$ also
for the case of spherical impenetrable particles. In real homogeneous composites with filler
shapes assimilable to spheres of diameter in the sub-micron range, the cross-over volume
fraction $\phi_c$ is consistently
larger than about $0.1$,\cite{Ambrosetti2009} so that a formula for $\delta_c/D$ that is useful
for real nanosphere composites must be accurate in the $\phi\gtrsim 0.1$ range.
For these $\phi$ values the scaling relation $\delta_c/D\propto \phi^{-1/3}$,
which stems by assuming very dilute systems such that $\delta_c/D\gg 1$, is of course no
longer valid. However, as noticed in Ref.~[\onlinecite{Johner2008}], the ratio
$\delta_c/\delta_{\rm NN}$, where $\delta_{\rm NN}$ is the mean minimal distance between
nearest-neighbours spheres, has a rather weak dependence on $\phi$. In particular,
we have found that the $\delta_c$ data for $a/b=1$ in Fig.~\ref{Fig4} are well
fitted by assuming that $\delta_c = 1.65\delta_{\rm NN}$ for $\phi\gtrsim 0.1$. An explicit formula
can then be obtained by using the high density asymptotic expression for $\delta_{\rm NN}$ as given in
Ref.~[\onlinecite{Torquato2001}]. This leads to:
\begin{equation}
\label{eq:deltasphere}
\delta_c/D\simeq \frac{1.65(1-\phi)^3}{12\phi(2-\phi)},
\end{equation}
which is plotted by the solid line in Fig.~\ref{Fig4}(a).

\section{analytic determination of the filler dependencies of the conductivity}
\label{formulas}

With the results of the previous section, we are now in a position to provide tunneling
conductivity formulas of random distributions of prolate, oblate and spherical objects for
$\sigma>\sigma_{\rm m}$, where $\sigma_{\rm m}$ is the intrinsic conductivity of the matrix. Indeed,
by substituting Eqs.~\eqref{eq:deltapro}, \eqref{eq:deltaOblates}, and \eqref{eq:deltasphere} into
Eq.~\eqref{eq:sigmaSP} we obtain
\begin{align}
\label{eq:sigmapro}
&\sigma\simeq\sigma_0\exp\!\left[-\frac{2D}{\xi}\frac{\gamma(b/a)^2}{\phi}\right] \, \, \,   \textrm{for prolates},  \\
\label{eq:sigmaobl}
&\sigma\simeq\sigma_0\exp\!\left\{-\frac{2D}{\xi}\left[\frac{0.15(a/b)}{\phi}\right]^{4/3}\right\} \, \, \, \textrm{for oblates}, \\
\label{eq:sigmasph}
&\sigma\simeq\sigma_0\exp\!\left[-\frac{2D}{\xi}\frac{1.65(1-\phi)^3}{12\phi(2-\phi)}\right] \, \, \, \textrm{for spheres} .
\end{align}
From the previously discussed conditions on the validity of the asymptotic formulas for $\delta_c/D$ it follows
that the above equations will hold when $(b/a)^2\lesssim\phi\lesssim 0.03$ for prolates, $\phi\gtrsim a/b$ and
$a/b < 0.1$ for oblates, and $\phi\gtrsim 0.1$  for spheres. We note in passing that for the case of
prolate objects, a relation of more general validity than Eq.~\eqref{eq:sigmapro} can be obtained by
substituting Eq.~\eqref{eq:deltapro2} into Eq.~\eqref{eq:sigmaSP}.

\begin{figure*}[t]
    \begin{center}
  \includegraphics[scale=0.3, clip='true']{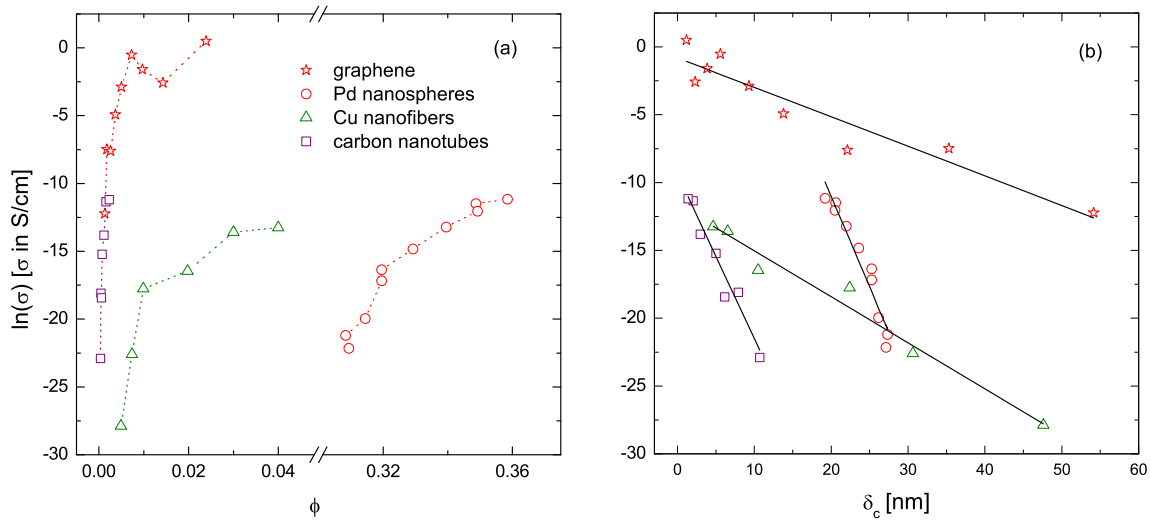}
  \caption{(Color online) (a):
  natural logarithm of the conductivity $\sigma$ as a function of the volume fraction $\phi$
  for different polymer nanocomposites: Graphene-polystyrene,\cite{Stankovich2006} Pd nanospheres-polystyrene,\cite{Kubat1993} Cu nanofibers-polystyrene,\cite{Gelves2006} and single-wall carbon nanotubes-epoxy.\cite{Bryning2005} When, for a given concentration, more then one value of $\sigma$ was
  given (as in Refs.~[\onlinecite{Gelves2006,Bryning2005}]), the average of $\ln (\sigma )$ was considered.
  (b), the same data of (a) re-plotted as function of the corresponding critical distance $\delta_c$.
  Solid lines are fits to Eq.~\eqref{eq:sigmaSPln}.}\label{Fig6}
  \end{center}
\end{figure*}

Although we are not aware of previous results on $\sigma$ for dispersions of oblate (plate-like) particles,
there exist nevertheless some results for prolate and spherical particles in the recent literature.
In Ref.~[\onlinecite{Hu2006}], for example, approximate expressions for $\sigma$ for extreme prolate ($a/b\gg 1$)
objects and their temperature dependence have been obtained by following the critical path method employed here.
It turns out that the temperature independent contribution to $\sigma$ that was given in  Ref.~[\onlinecite{Hu2006}]
has the same dependence on the particle geometry and density of Eq.~\eqref{eq:sigmapro}, but without the
numerical coefficients. The case of relatively high density spheres has been considered in
Ref.~[\onlinecite{Balberg2009}] where $\ln(\sigma)\propto 1/\phi$ has been proposed. This implies that
$\delta_c/D\propto 1/\phi$, which does not adequately fit the numerical results of $\delta_c/D$, while
Eq.~\eqref{eq:deltasphere}, and consequently Eq.~\eqref{eq:sigmasph}, are rather accurate for a wide
range of $\phi$ values.

In addition to the $\phi$-dependence of the tunneling contribution to the conductivity,
Eqs.~\eqref{eq:sigmapro}-\eqref{eq:sigmasph} provide also estimations for the cross-over value $\phi_c$,
below which the conductivity basically coincides with the conductivity $\sigma_{\rm m}$ of the insulating matrix.
As discussed in Sec.\ref{cond}, and as illustrated in Fig.~\ref{Fig3}, $\phi_c$ may be estimated
by the $\phi$ value such that $\sigma\simeq\sigma_{\rm m}$, which leads to
\begin{equation}
\label{eq:phicpro}
\phi_c\simeq\frac{2D}{\xi}\frac{\gamma (b/a)^2}{\ln(\sigma_0/\sigma_{\rm m})},
\end{equation}
for prolate and
\begin{equation}
\label{eq:phicobl}
\phi_c\simeq 0.15 (a/b)\left[\frac{2D}{\xi}\frac{1}{\ln (\sigma_0/\sigma_{\rm m})}\right]^{3/4},
\end{equation}
for oblate objects. For the case of spheres, $\phi_c$ is the root of a third-order
polynomial equation. Equations \eqref{eq:phicpro} and \eqref{eq:phicobl}, by construction,
display the same dependence on the aspect-ratio
of the corresponding geometrical percolation critical densities, as it can be appreciated by comparing
them with Eq.~\eqref{eq:deltapro} (prolates) or with the inverse of Eq.~\eqref{eq:deltaOblates} (oblates).
However they also show that the cross-over point depends on the tunneling decay length and on the
intrinsic matrix conductivity. This implies that if, by some means, one could alter $\sigma_{\rm m}$ in a given
composites without seriously affecting $\xi$ and $\sigma_0$, then a change in $\phi_c$ is to be expected.

\section{comparison with experimental data}
\label{compa}

\begin{figure*}[t]
    \begin{center}
  \includegraphics[scale=0.35, clip='true']{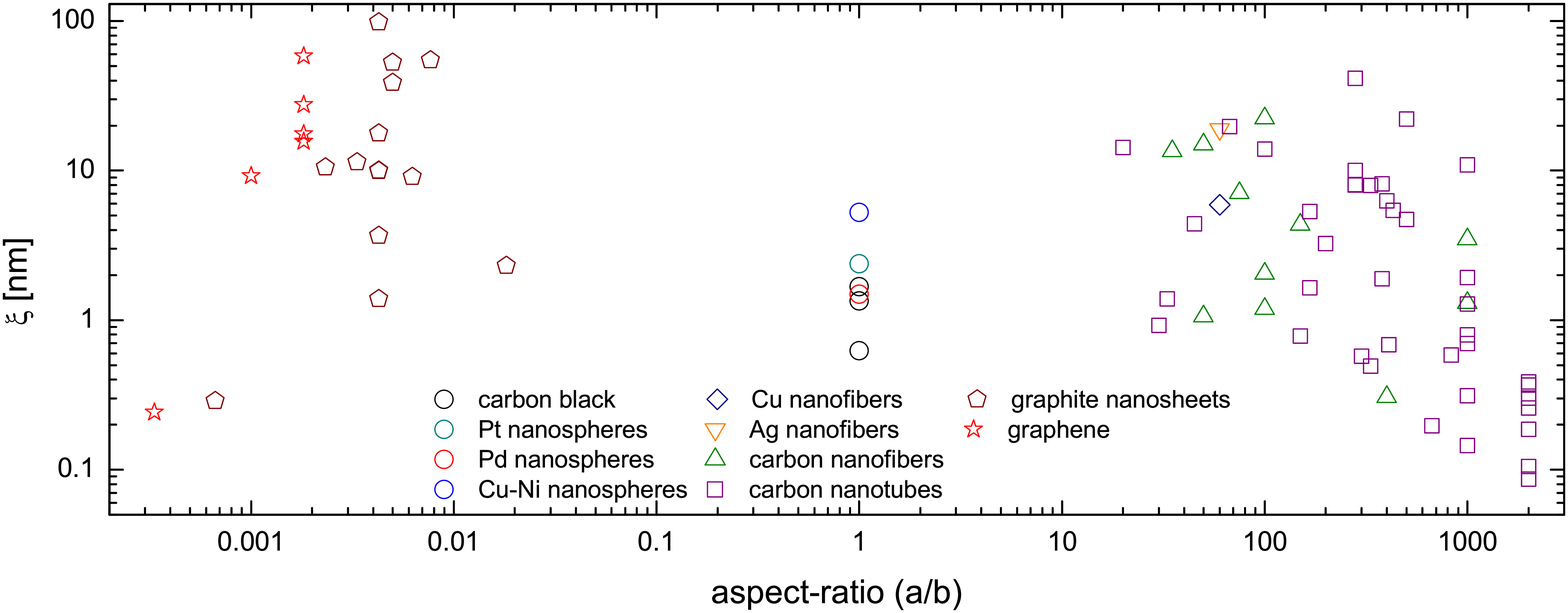}
  \caption{(Color online)
  Characteristic tunneling distance $\xi$ values for different polymer nanocomposites as extracted form
  Eq.~\eqref{eq:sigmaSPln} applied to the data of Refs.~[\onlinecite{Kubat1993,nanosph}]
  (low structured carbon black and metallic nanosphere composites),
  Refs.~[\onlinecite{Gelves2006,Bryning2005,nanopro}] (nanofiber and carbon nanotube composites), and
  Refs.~[\onlinecite{Stankovich2006,Lu2006,nanoobl}] (nanographite and graphene composites).}\label{Fig7}
  \end{center}
\end{figure*}

In this section we show how the above outlined formalism may be used to re-interpret the
experimental data on the conductivity of different nanocomposites that were reported in the literature.
In Fig.~\ref{Fig6}(a) we show the measured data of $\ln(\sigma)$
versus $\phi$ for polymer composites filled with graphene sheets,\cite{Stankovich2006}
Pd nanospheres,\cite{Kubat1993} Cu nanofibers,\cite{Gelves2006} and carbon nanotubes.\cite{Bryning2005}
Equation \eqref{eq:sigmaSP} implies that the same
data can be profitably re-plotted as a function of $\delta_c$, instead of  $\phi$. Indeed, from
\begin{equation}
\label{eq:sigmaSPln}
\ln(\sigma)=-\frac{2}{\xi}\delta_c+\ln(\sigma_{0}),
\end{equation}
we expect a linear behavior, with a slope $-2/\xi$, that is independent of the specific value of $\sigma_0$,
which allows for a direct evaluation of the characteristic tunneling distance $\xi$.
By using the values of $D$ and $a/b$ provided
in Refs.~[\onlinecite{Stankovich2006,Kubat1993,Gelves2006,Bryning2005}] (see also Appendix B)
and Eqs.~\eqref{eq:deltapro2}, \eqref{eq:deltaOblates}, \eqref{eq:deltasphere} for $\delta_c$, we
find indeed an approximate linear dependence on $\delta_c$ [Fig.~\ref{Fig6}(b)], from which we
extract $\xi\simeq 9.22$ nm for graphene, $1.50$ nm for the nanospheres, $5.9$ nm for the nanofibers,
and $1.65$ nm for the nanotubes.

We further applied this procedure to several published data on polymer-based composites with
nanofibers and carbon nanotubes,\cite{nanopro} nanospheres,\cite{nanosph} and
nanosheets (graphite and graphene),\cite{Lu2006,nanoobl} hence with fillers
having $a/b$ ranging from $\sim 10^{-3}$ up to $\sim 10^{3}$.
As detailed in Appendix B, we have fitted Eq.~\eqref{eq:sigmaSPln} to the
experimental data by using our formulas for $\delta_c$.
The results are collected in Fig.~\ref{Fig7}, showing that most of the so-obtained values of the tunneling
length $\xi$ are comprised between $\sim 0.1$ nm and $\sim 10$ nm, in accord with the expected
value range.\cite{Balberg1987b,Shklovskii1984,Holmlin2001,Seager1974,Benoit2002}
This is a striking result considering the number of factors that make a real composite deviate from an
idealized model. Most notably, fillers may have non-uniform size, aspect-ratio, and geometry,
they may be oriented, bent and/or coiled, and interactions with the polymer may lead to agglomeration,
segregation, and sedimentation.
Furthermore, composite processing can alter the properties of the pristine fillers,
e.g. nanotube or nanofiber breaking (which may explain the downward drift of $\xi$ for high aspect-ratios
in Fig.~\ref{Fig7}) or graphite nanosheet exfoliation (which may explain the upward shift of $\xi$
for the graphite data). In principle, deviations from ideality can be included in the present formalism
by evaluating their effect on $\delta_c$.\cite{Kyrylyuk2008} It is however interesting to notice that
all these factors have often competing effects in raising or lowering the composite conductivity, and
Fig.~\ref{Fig6} suggests that on the average they compensate each other to some extent, allowing
tunneling conduction to strongly emerge from the $\phi$-dependence of $\sigma$ as a visible characteristic
of nanocomposites.

\section{discussion and conclusions}
\label{concl}

As discussed in the introduction, the theory of conductivity in nanocomposites presented in the previous
sections is based on the observation that a microscopic mechanism of interparticle conduction based
on tunneling is not characterized by any sharp cut-off, so that the composite conductivity is not expected
to follow the percolation-like behavior of Eq.~\eqref{plaw}. Nevertheless, we have demonstrated that
concepts and quantities pertinent to percolation theory, like the critical path approximation and the
associated critical path distance $\delta_c$, are very effective in
describing tunneling conductivity in composite materials.
In particular, we have shown that the (geometrical) connectivity
problem of semi-penetrable objects in the continuum, as discussed in Sec.~\ref{critical}, is of
fundamental importance for the understanding of the filler dependencies ($\phi$, $D$, and $a/b$)
of $\sigma$, and that it gives the possibility to formulate analytically such dependencies, at least
for some asymptotic regimes. In this respect, the body of work which can be found on the connectivity
problem in the literature finds a straightforward applicability in the present context of transport in
nanocomposites. For example, it is not uncommon to find studies on the connectivity
of semi-penetrable objects in the continuum where the thickness $\delta/2$ of the penetrable shell is
phenomenologically interpreted as a distance of the order of the tunneling length
$\xi$.\cite{Berhan2007,Kyrylyuk2008,Ambrozic2005,Hicks2009} This interpretation is replaced here by
Eq.~\eqref{eq:sigmaSP} which provides a clear recipe for the correct use, in the context of
tunneling, of the connectivity problem through the critical thickness $\delta_c/2$.
Furthermore, Eq.~\eqref{eq:sigmaSP} could be applied to nanocomposite systems where, in addition to the
hard-core repulsion between the impenetrable particles, effective inter-particle interactions are
important, such as those arising from depletion interaction in polymer-base composites. In this respect,
recent theoretical results on the connectivity of polymer-nanotube composites may find a broader
applicability in the present context.\cite{Kyrylyuk2008}

It is also worth noticing that, although our results on the filler dependencies of $\delta_c$
for prolate objects with $a/b\gg 1$ can be understood from the consideration of excluded volume effects (e.g. second
virial approximation), the corresponding $\delta_c$ formulas for the oblate and spherical cases
are empirical, albeit rather accurate with respect to our Monte Carlo results. It would be therefore
interesting to find microscopic justifications to our results, especially for the case of oblates
with $a/b\ll 1$, which appear to display a power-law dependence of $\delta_c$ on the volume fraction
[Eq.~\eqref{eq:deltaOblates}].
\begin{figure}[t]
    \begin{center}
  \includegraphics[scale=0.3, clip='true']{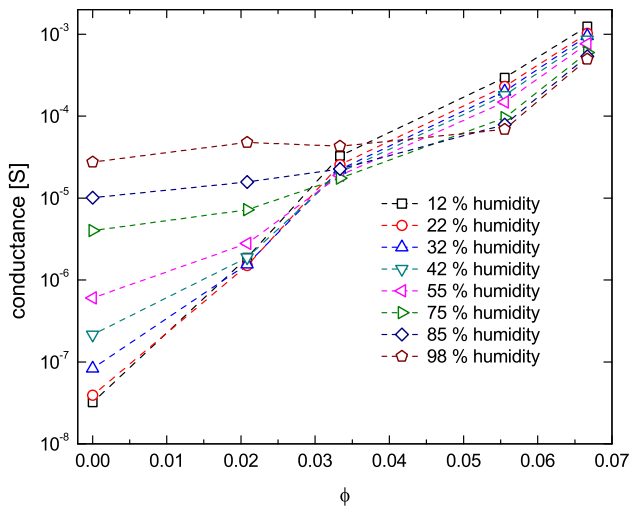}
  \caption{(Color online)
  Conductance versus $\phi$ dependence for a carbon black-quaternized Poly(4-vinylpyridine) composite
  for different humidities. Adapted from Ref.~[\onlinecite{Li2007b}]}\label{Fig8}
  \end{center}
\end{figure}

Let us now turn to discuss some consequences of the theory presented here. As shown in Sec.~\ref{formulas}
the cross-over volume fraction $\phi_c$ depends explicitly on the conductivity $\sigma_{\rm m}$ of the
insulating medium, leading to the possibility of shifting $\phi_c$ by altering $\sigma_{\rm m}$.
Formulas \eqref{eq:phicpro} and \eqref{eq:phicobl} were obtained by assuming that the transport mechanism
leading to $\sigma_{\rm m}$ was independent of the concentration of the conducting fillers, as it is the case for
polymer-based nanocomposites, where the conduction within the polymer is due to ion mobility. In that case,
a change in $\sigma_{\rm m}$, and so a change in $\phi_c$,  could be induced by a change in the ion concentration.
This is nicely illustrated by an example where a conductive polymer composite with large ionic conductivity
was studied as a material for humidity sensors.\cite{Li2007b}
This consisted of carbon black dispersed in a Poly(4-vinylpyridine) matrix which was quaternized in order
to obtain a polyelectrolyte. Since the absorbed water molecules interact with the polyelectrolyte and
facilitate the ionic dissociation, higher humidity implies a larger ionic conductivity.
In Fig.~\ref{Fig8} we have redrawn Fig. 4 of Ref.~[\onlinecite{Li2007b}] in terms of the conductance as a
function of carbon black content for different humidity levels. Consistently with our assumptions,
one can see that with the increase of humidity the matrix intrinsic conductivity is indeed shifted upward,
while this has a weaker effect on the conductivity for higher contents of carbon black, where transport
is governed by interparticle tunneling (a slight downshift in this region is attributed to enhanced
interparticle distances due to water absorption). The net effect illustrated in Fig.~\ref{Fig8} is thus
a shift of the crossover point $\phi_c$ towards higher values of carbon black content. It is worth
noticing that the explanation proposed by the authors of Ref.~[\onlinecite{Li2007b}] in order to account
for their finding is equivalent to the global tunneling network/crossover scenario.

Another feature which should be expected by the global tunneling network model concerns
the response of the conductivity to an applied strain $\varepsilon$. Indeed, by using
Eq.~\eqref{eq:sigmaSP}, the piezoresistive response $\Gamma$,
that is the relative change of the resistivity $\sigma^{-1}$ upon an applied $\varepsilon$,
reduces to:
\begin{equation}
\label{piezo}
\Gamma\equiv \frac{d\ln(\sigma^{-1})}{d\varepsilon}=\ln\!\left(\frac{\sigma_0}{\sigma}\right)
\frac{d\ln(\delta_c)}{d\varepsilon}.
\end{equation}
In the above expression $d\ln(\delta_c)/d\varepsilon=1$ for fillers having the same elastic properties
of the insulating matrix. In contrast, for elastically rigid fillers this term
can be rewritten as $[d\ln(\delta_c)/d\ln(\phi)]d\ln(\phi)/d\varepsilon$, which is also approximatively
a constant due to the $\delta_c$ dependence on $\phi$ as given in Eqs.~\eqref{eq:deltapro},
\eqref{eq:deltaOblates}, and \eqref{eq:deltasphere}, and to $d\ln(\phi)/d\varepsilon\simeq -1$.
Hence, the expected dominant dependence of $\Gamma$ is of the form $\Gamma\propto \ln(1/\sigma)$, which
has been observed indeed in Refs.[\onlinecite{Tamborin1997,Niki2009}].

Finally, before concluding, we would like to point out that, with the theory presented in this paper,
both the low temperature and the filler dependencies of nanocomposites in the dielectric regime
have a unified theoretical framework. Indeed, by taking into account particle excitation energies,
Eq.~\eqref{eq:tunneling} can be generalized to include inter-particle electronic
interactions, leading, within the critical path approximation, to a critical distance $\delta_c$ which
depends also on such interactions and on the temperature. The resulting generalized theory would
be equivalent then to the hopping transport theory
corrected by the excluded volume effects of the impenetrable cores of the conducting particles.
An example of this generalization for the case of nanotube composites is the work of Ref.~[\onlinecite{Hu2006}].

In summary, we have considered the tunneling-percolation problem in the so far unstudied intermediate
regime between the percolation-like and the hopping-like regimes by extending
the critical path analysis to systems and properties that are pertinent to nanocomposites.
We have analyzed published conductivity data for several nanotubes, nanofibers, nanosheets,
and nanospheres composites and extracted
the corresponding values of the tunneling decay length $\xi$.
Remarkably, most of the extracted $\xi$ values fall within its expected range, showing that tunneling is
a manifested characteristic of the conductivity of nanocomposites.
Our formalism can be used to tailor the electrical properties of real composites, and can be generalized
to include different filler shapes, filler size and/or aspect-ratio polydispersity, and
interactions with the insulating matrix.

\acknowledgements
This study was supported in part by the Swiss Commission for Technological Innovation (CTI) through
project GraPoly, (CTI Grant No. 8597.2), a joint collaboration led by TIMCAL Graphite \& Carbon SA,
in part by the Swiss National Science Foundation (Grant No. 200021-121740), and in part by the Israel
Science Foundation (ISF). Discussions with E. Grivei and N. Johner are greatly appreciated.

\appendix

\section{Excluded volumes of spheroids and spherocylinders with isotropic orientation distribution}

The work of Isihara \cite{Isihara1950} enables to derive closed relations for the excluded volume of two
spheroids with a shell of constant thickness and for
an isotropic distribution of the mutual orientation of the spheroid symmetry axes.
Given two spheroids with polar semi-axis $a$ and equatorial semi-axis $b$, their eccentricity
$\epsilon$ are defined as follows:
\begin{align}
\epsilon=&\sqrt{1-\frac{b^{2}}{a^{2}}}\qquad\qquad\textrm{for prolates}, \\
\epsilon=&\sqrt{1-\frac{a^{2}}{b^{2}}}\qquad\qquad\textrm{for oblates}.
\end{align}
If the mutual orientation of the spheroid symmetry axes is isotropic, the averaged
excluded volume of the two spheroids is then (valid also for more general
identical ovaloids)
\begin{equation}
\label{eq:Vex_gen}
V_{\rm exc}=2V+\frac{MF}{2\pi},
\end{equation}
where $V$ is the spheroid volume and $M$ and $F$ are two quantities defined as:\cite{Isihara1950}
\begin{align}
M=&2\pi a
\bigg[1+\frac{(1-\epsilon^{2})}{2\epsilon}
\ln{\bigg(\frac{1+\epsilon}{1-\epsilon}\bigg)}\bigg], \\
F=&2\pi ab \bigg(\sqrt{1-\epsilon^{2}}+\frac{\arcsin{\epsilon}}{\epsilon}\bigg),
\end{align}
for the case of prolate ($a/b>1$) spheroids and
\begin{align}
M=&2\pi b\bigg(\sqrt{1-\epsilon^{2}}+\frac{\arcsin{\epsilon}}{\epsilon}\bigg), \\
F=&2\pi b^{2} \bigg[ 1+\frac{(1-\epsilon^{2})}{2\epsilon}
\ln{\bigg(\frac{1+\epsilon}{1-\epsilon}\bigg)}\bigg],
\end{align}
for the case of oblate ($a/b<1$) spheroids.

If now the spheroids are coated with a shell of uniform thickness $d$ ($d=\delta/2$),
then the averaged excluded volume of the spheroids plus shell has again the form of (\ref{eq:Vex_gen}):
\begin{equation}
 V_{\rm exc}^{\rm tot}=2V_{d}+\frac{M_{d}F_{d}}{2\pi},
\end{equation}
and by constructing the quantities $V_{d}$, $M_{d}$, and $F_{d}$ from their definition in
Ref.~[\onlinecite{Isihara1950}] (see Ref.~[\onlinecite{Ambrosetti2008}] for a similar calculation),
one obtains:
\begin{equation}
\label{eq:Vexdgeneral}
V_{\rm exc}^{\rm tot}= V_{\rm exc} + 4dF + \frac{dM^{2}}{\pi} + 8d^{2}M + \frac{32\pi}{3}d^{3}.
\end{equation}
In the cases of extreme prolate ($a/b\gg 1$ and $\delta/a\ll 1$) and oblate
($a/b\ll 1$ and $\delta/b\ll 1$) spheroids, the total excluded
volume reduces therefore to
\begin{equation}
\label{excpro}
V_{\rm exc}^{\rm tot}=V_{\rm exc}+2\pi a^2 \delta,
\end{equation}
for prolates and
\begin{equation}
V_{\rm exc}^{\rm tot}=V_{\rm exc}+4\pi b^2\delta+\pi^3 b^2\delta/2+2\pi^2 b\delta^2,
\end{equation}
for oblates.
Within the second-order virial approximation, the critical distance $\delta_c$ is related to the
volume fraction $\phi$ through $\phi\simeq V/\Delta V_{\rm exc}$,
where $\Delta V_{\rm exc}=V_{\rm exc}^{\rm tot}-V_{\rm exc}$. From the above expressions one has then
[$D=2\max (a,b)$]:
\begin{equation}
\label{phipro}
\phi\simeq \frac{(b/a)^2}{3\delta_c /D},
\end{equation}
for prolates and
\begin{equation}
\label{phiobl}
\phi\simeq \frac{(4/3)(a/b)}{(8+\pi^2)\delta_c/D+8\pi(\delta_c/D)^2},
\end{equation}
for oblates.

For comparison, we provide below the excluded volumes of randomly oriented
spherocylinders. These are formed by cylinders of radius $R$ and length $L$, capped by hemispheres of radius $R$.
Their volume is  $V=(4/3)\pi R^3[1+(3/4)(L/R)]$. The excluded volume for spherocylinders with isotropic orientation distribution was calculated in Ref.~[\onlinecite{Balberg1984}] and reads
\begin{equation}
\label{eq:VexSpherocyl}
V_{\rm exc}=\frac{32 \pi}{3} R^3 \left[1+\frac{3}{4}(L/R)+\frac{3}{32}(L/R)^2\right].
\end{equation}
The excluded volume $V_{\rm exc}^{\rm tot}$ of spherocylinders with a shell of constant thickness
$d=\delta/2$ is hen:
\begin{equation}
\label{eq:VexdSpherocyl}
V_{\rm exc}^{\rm tot}=\frac{32 \pi}{3} (R+d)^3 \left[1+\frac{3}{4}\left(\frac{L}{R+d}\right)+\frac{3}{32}\left(\frac{L}{R+d}\right)^2\right].
\end{equation}
For the high aspect-ratio limit ($L/R\gg 1$), when $d/L\ll 1$, the total excluded volume minus the excluded
volume of the impenetrable core is
\begin{equation}
\Delta V_{\rm exc}=\pi L^2 d,
\end{equation}
which coincides with the last term of Eq.~\eqref{excpro} if $d=\delta/2$ and $a=R+L/2\simeq L/2$.
Furthermore, the second-order virial approximation ($\phi\simeq V/\Delta V_{\rm exc}$) gives:
\begin{equation}
\phi\simeq \frac{(b/a)^2}{2\delta_c/D},
\end{equation}
which has a numerical coefficient different from Eq.~\eqref{phipro} because for spherocylinders
$V\simeq 2\pi a b^2$ (for $a/b\gg 1$) while for spheroids $V=4\pi a b^2/3$.

\newpage
\widetext

\section{Supplementary material. Conductivity versus critical distance plots}

We show in the following Figs.~\ref{fig:SMfig4}-\ref{fig:SMfig12} the complete set of plots of the natural logarithm of
the sample conductivity $\sigma$ as a function of the geometrical percolation critical distance $\delta_c$ for
different polymer nanocomposites, as used to obtain the $\xi$ values of Fig.~7 of the main article.
In collecting the published results of $\sigma$ versus $\phi$, we have considered only those works where
$a/b$ and $D=2\max(a,b)$ were explicitly reported. In the cases of documented variations of these quantities,
we used their arithmetic mean.
The $\phi$ dependence of the original published data was then converted into a $\delta_c$ dependence as follows.
For fibrous systems (nanofibers, nanotubes), the filler shape was assimilated to spherocylinders, while for
nanosheet systems it was assimilated to oblate spheroids.
For prolate fillers  $\delta_c$ was obtained from Eq.~(5), for oblate fillers
the values for $\delta_c$ were obtained from Eq.~(8), while
for spherical fillers, the values of $\delta_c$ were obtained from Eq.~(9).

Since the model introduced in the main text is expected to be representative only if $\phi$ is sufficiently above $\phi_c$ to consider the effect of the insulating matrix negligible, for a given experimental curve, higher $\phi$ data were privileged, and lower density points sometimes omitted when deviating consistently from the main trend.
The converted data were fitted to Eq.~(15) of the main text and the results of the fit are reported in Figs.~\ref{fig:SMfig4}-\ref{fig:SMfig12} by solid lines. The results for $2/\xi$ and
$\ln(\sigma_0)$ are also reported in the figures.  As it may be appreciated from Figs.~\ref{fig:SMfig4}-\ref{fig:SMfig12}, in many instances the experimental data follow
nicely a straight line, as predicted by Eq.~(15) of the main text, while in others the data are rather scattered or deviate from linearity. In these latter cases, the fit to Eq.~(15)
is meant to capture the main linear trend of  $\ln(\sigma)$ as a function of $\delta_c$. It should also be noticed that, in spite of the rather narrow distribution of the extracted
$\xi$ values reported in Fig.~7 of the main text, the values of the prefactor $\sigma_0$ obtained from the
fits are widely dispersed. This is of course due to the fact that, besides intrinsic
variations of the tunneling prefactor conductance for different composites,
interpolating the data to $\delta_c=0$ leads to a large variance  of $\sigma_0$ even for
minute changes of the slope.  We did not notice any significant correlation between the extracted
$\xi$ and $\sigma_0$ values.

\begin{figure}[h!]
    \begin{center}
  \includegraphics[scale=1.0, clip='true']{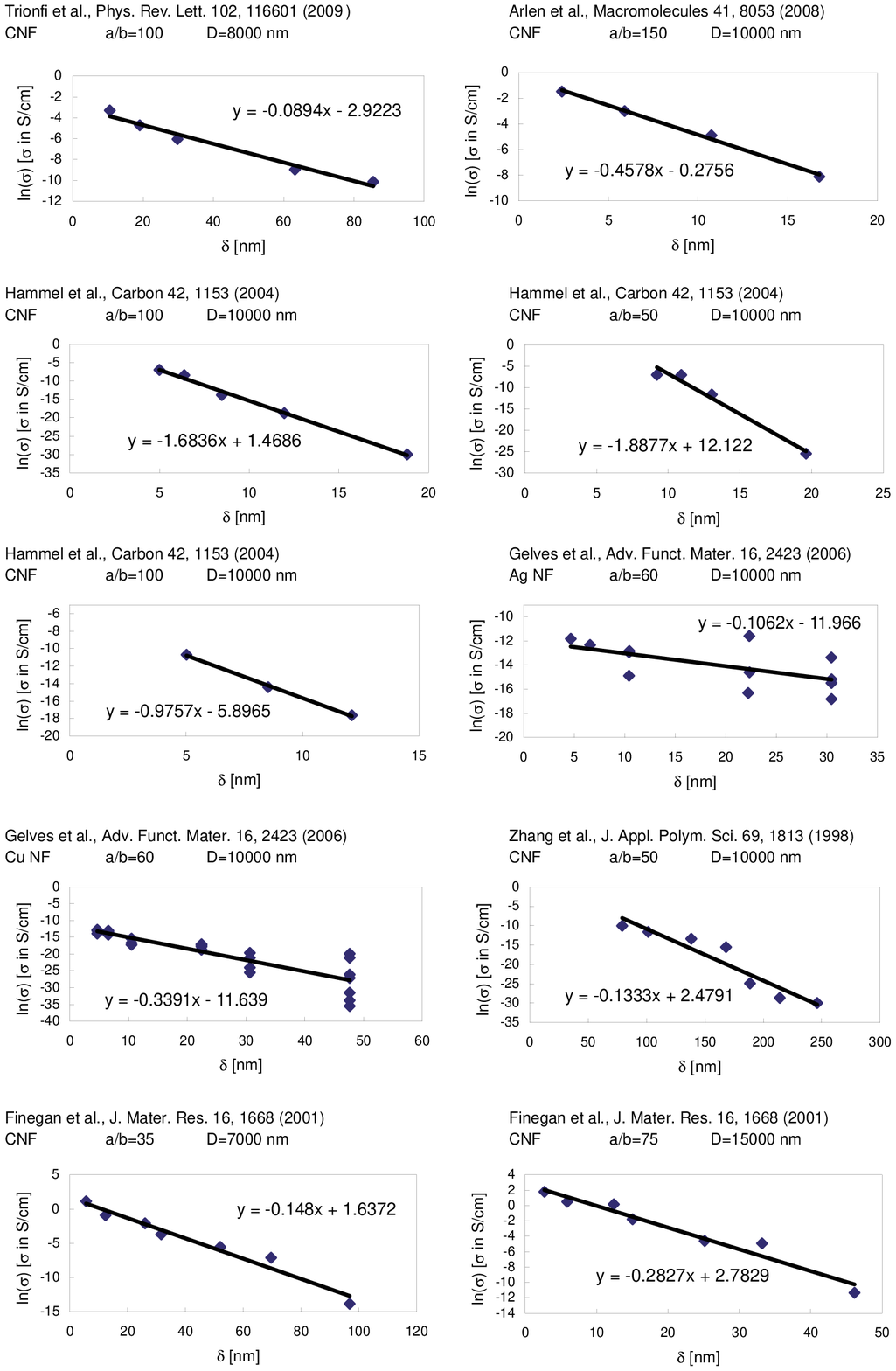}
  \caption{plots of $\ln\sigma$ as a function of $\delta$ for different polymer-nanofiber composites.}\label{fig:SMfig4}
  \end{center}
\end{figure}

\begin{figure}[h!]
    \begin{center}
  \includegraphics[scale=1.0, clip='true']{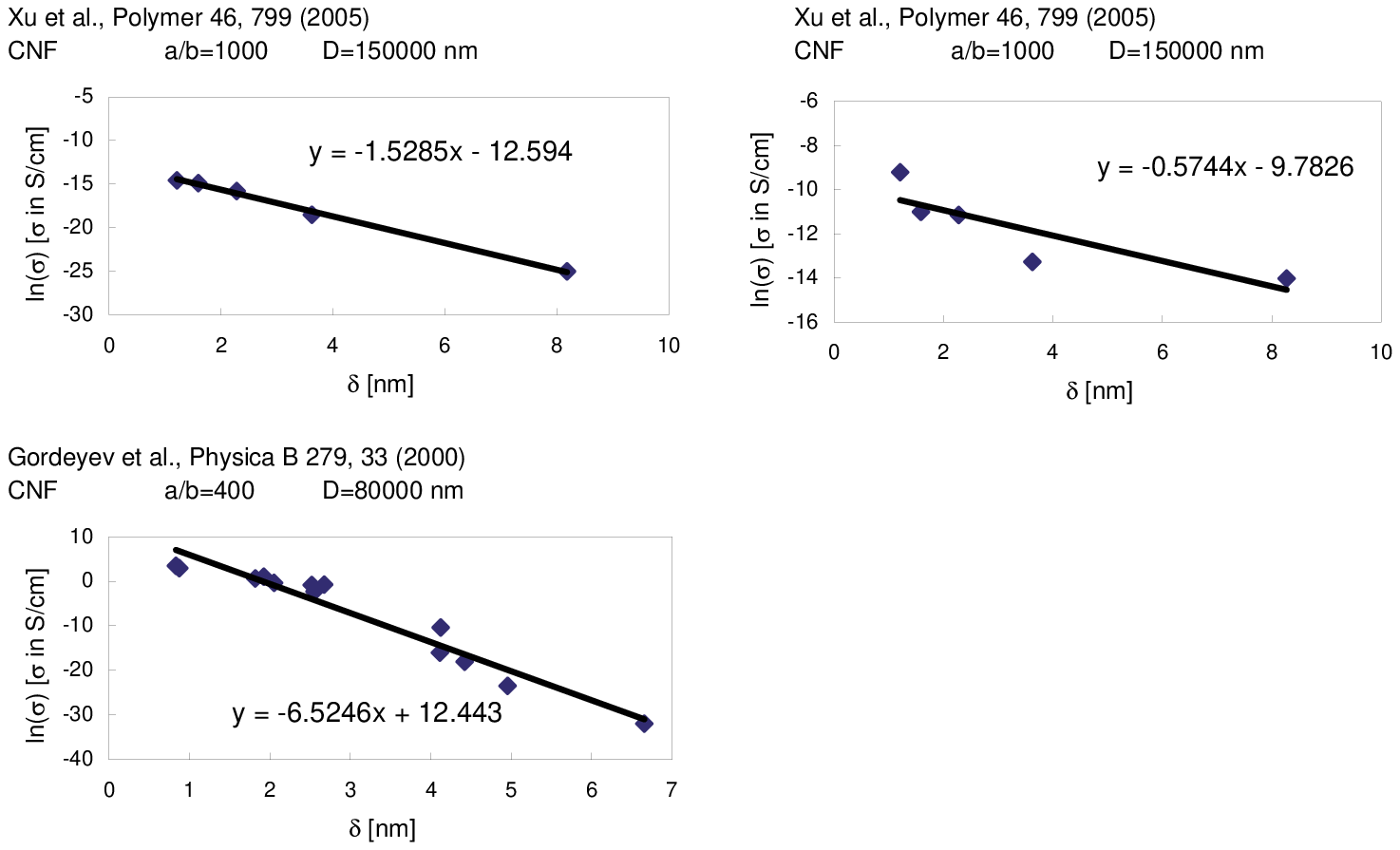}
  \caption{plots of $\ln\sigma$ as a function of $\delta$ for different polymer-nanofiber composites (cont.)}\label{fig:SMfig5}
  \end{center}
\end{figure}

\begin{figure}[h!]
    \begin{center}
  \includegraphics[scale=1.0, clip='true']{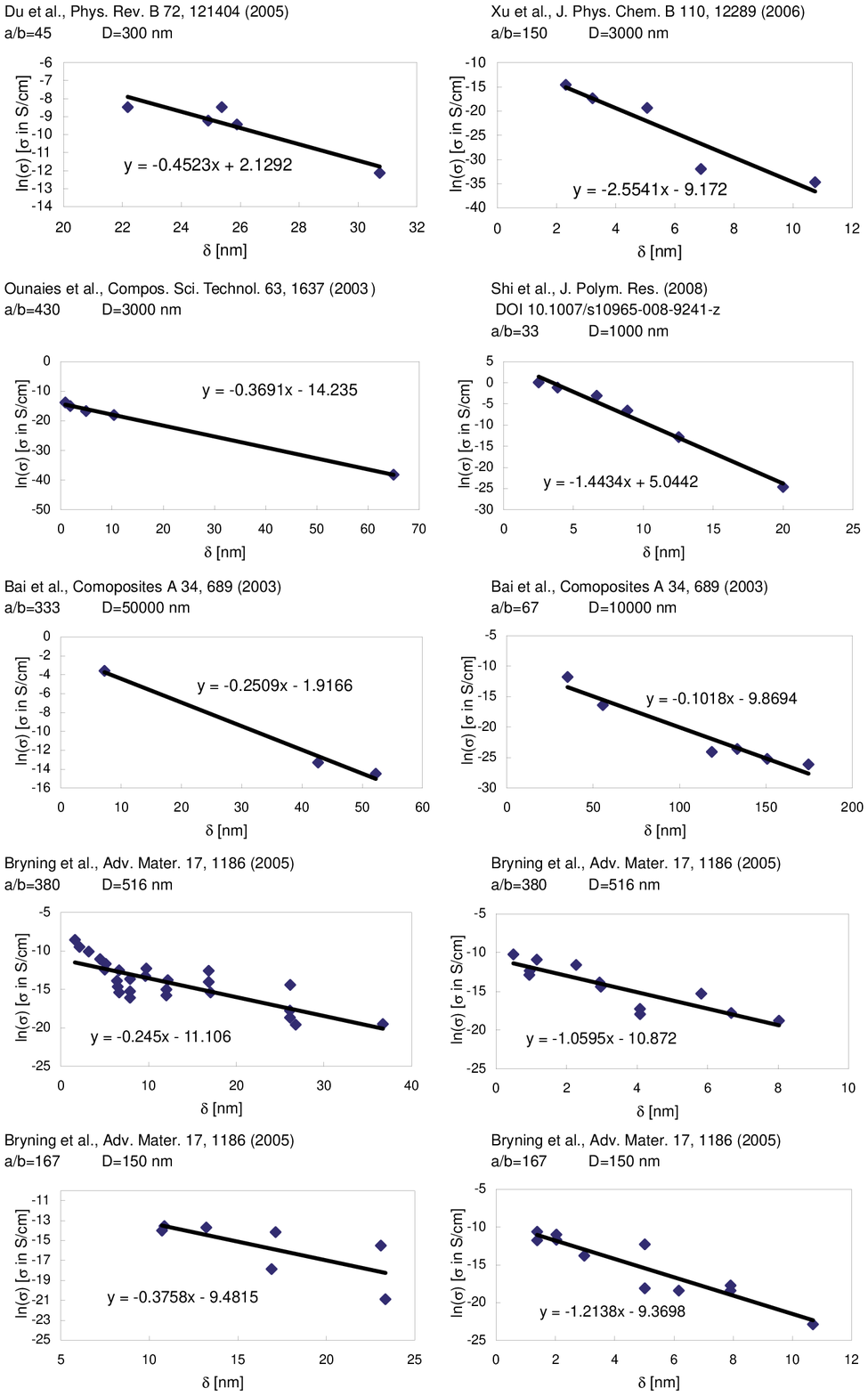}
  \caption{plots of $\ln\sigma$ as a function of $\delta$ for different polymer-nanotube composites.}\label{fig:SMfig6}
  \end{center}
\end{figure}

\begin{figure}[h!]
    \begin{center}
  \includegraphics[scale=1.0, clip='true']{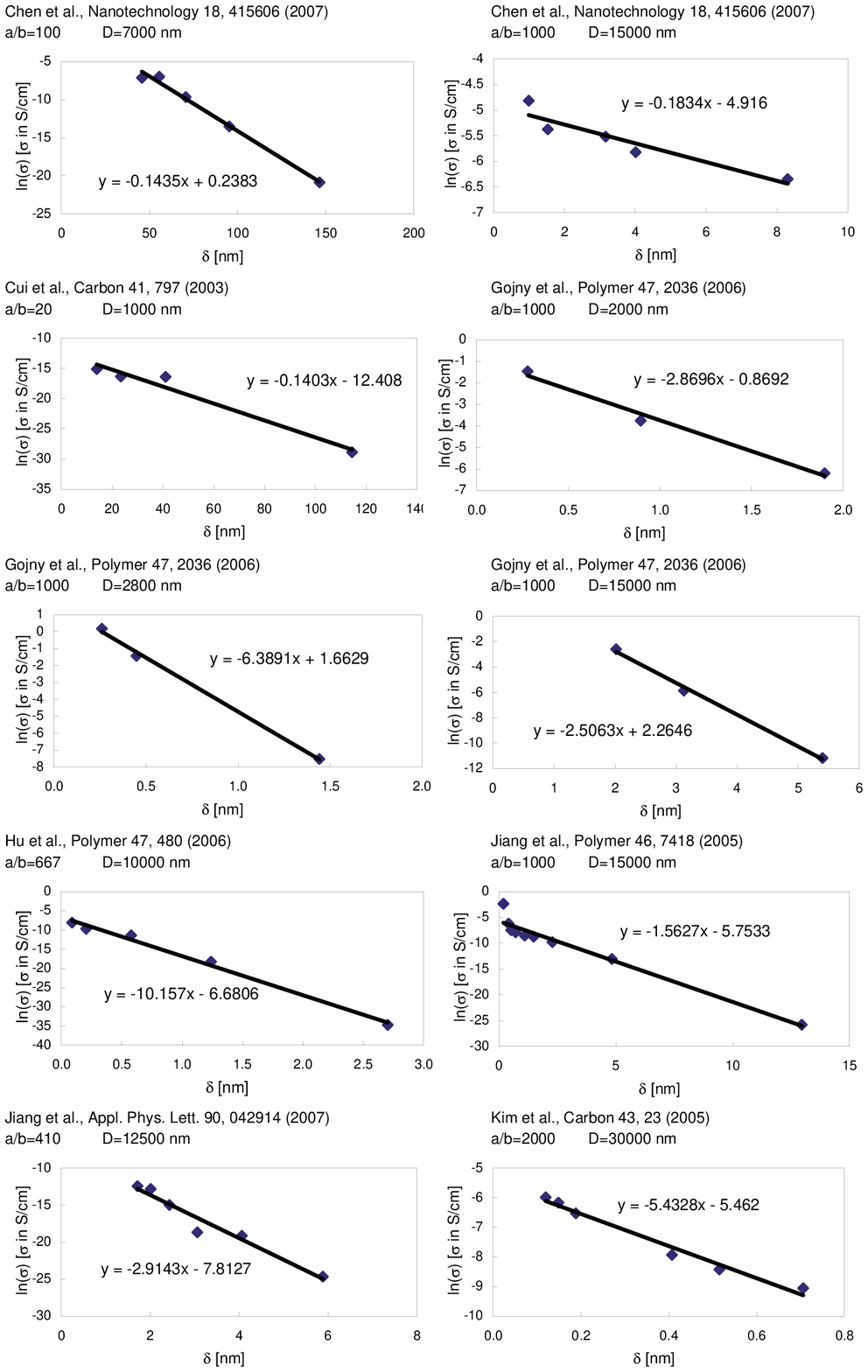}
  \caption{plots of $\ln\sigma$ as a function of $\delta$ for different polymer-nanotube composites (cont.)}\label{fig:SMfig7}
  \end{center}
\end{figure}

\begin{figure}[h!]
    \begin{center}
  \includegraphics[scale=1.0, clip='true']{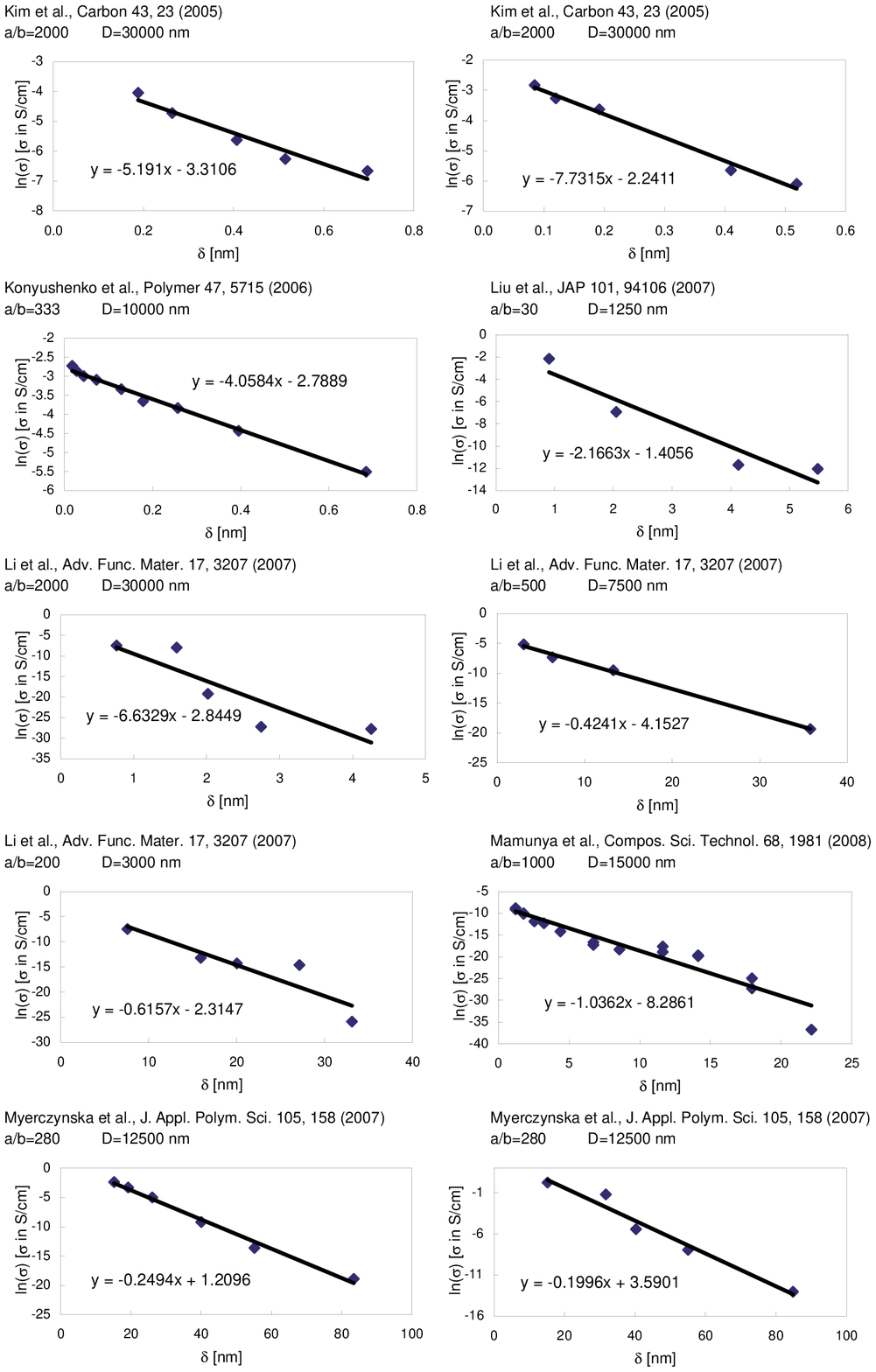}
  \caption{plots of $\ln\sigma$ as a function of $\delta$ for different polymer-nanotube composites (cont.)}\label{fig:SMfig8}
  \end{center}
\end{figure}

\begin{figure}[h!]
    \begin{center}
  \includegraphics[scale=1.0, clip='true']{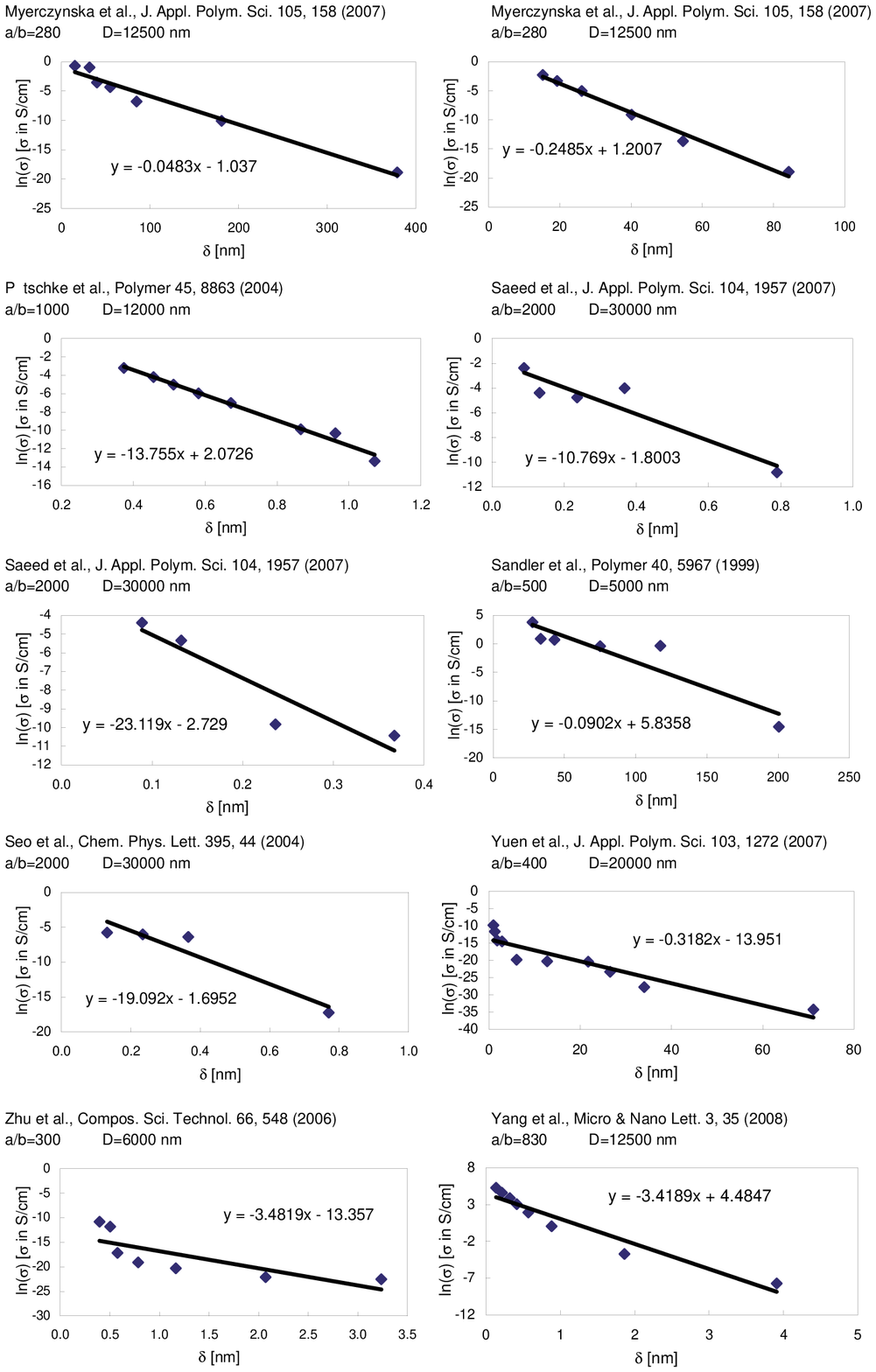}
  \caption{plots of $\ln\sigma$ as a function of $\delta$ for different polymer-nanotube composites (cont.)}\label{fig:SMfig9}
  \end{center}
\end{figure}

\begin{figure}[h!]
    \begin{center}
  \includegraphics[scale=1.0, clip='true']{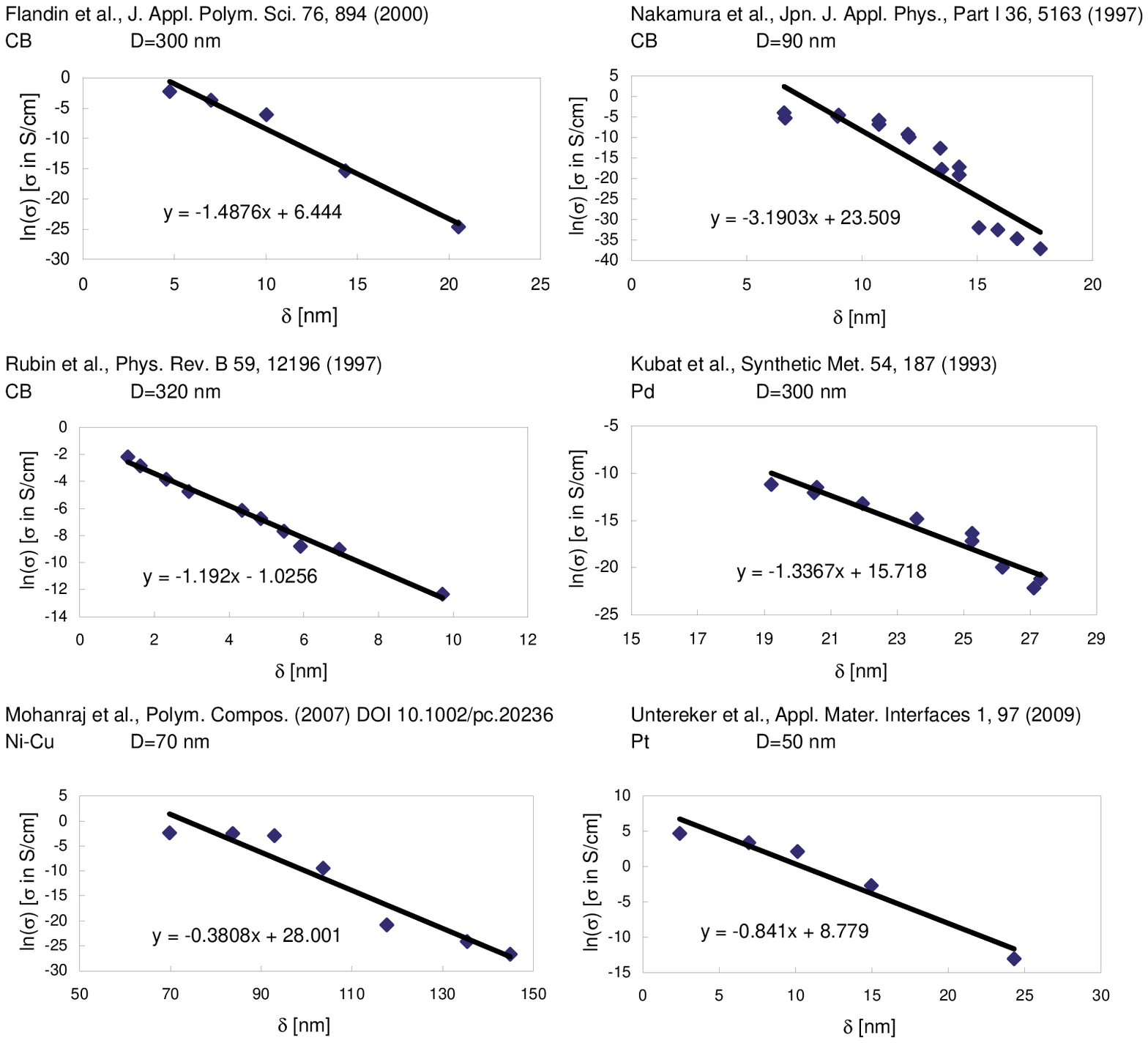}
  \caption{plots of $\ln\sigma$ as a function of $\delta$ for different polymer-nanospheres composites }\label{fig:SMfig10}
  \end{center}
\end{figure}

\begin{figure}[h!]
    \begin{center}
  \includegraphics[scale=1.0, clip='true']{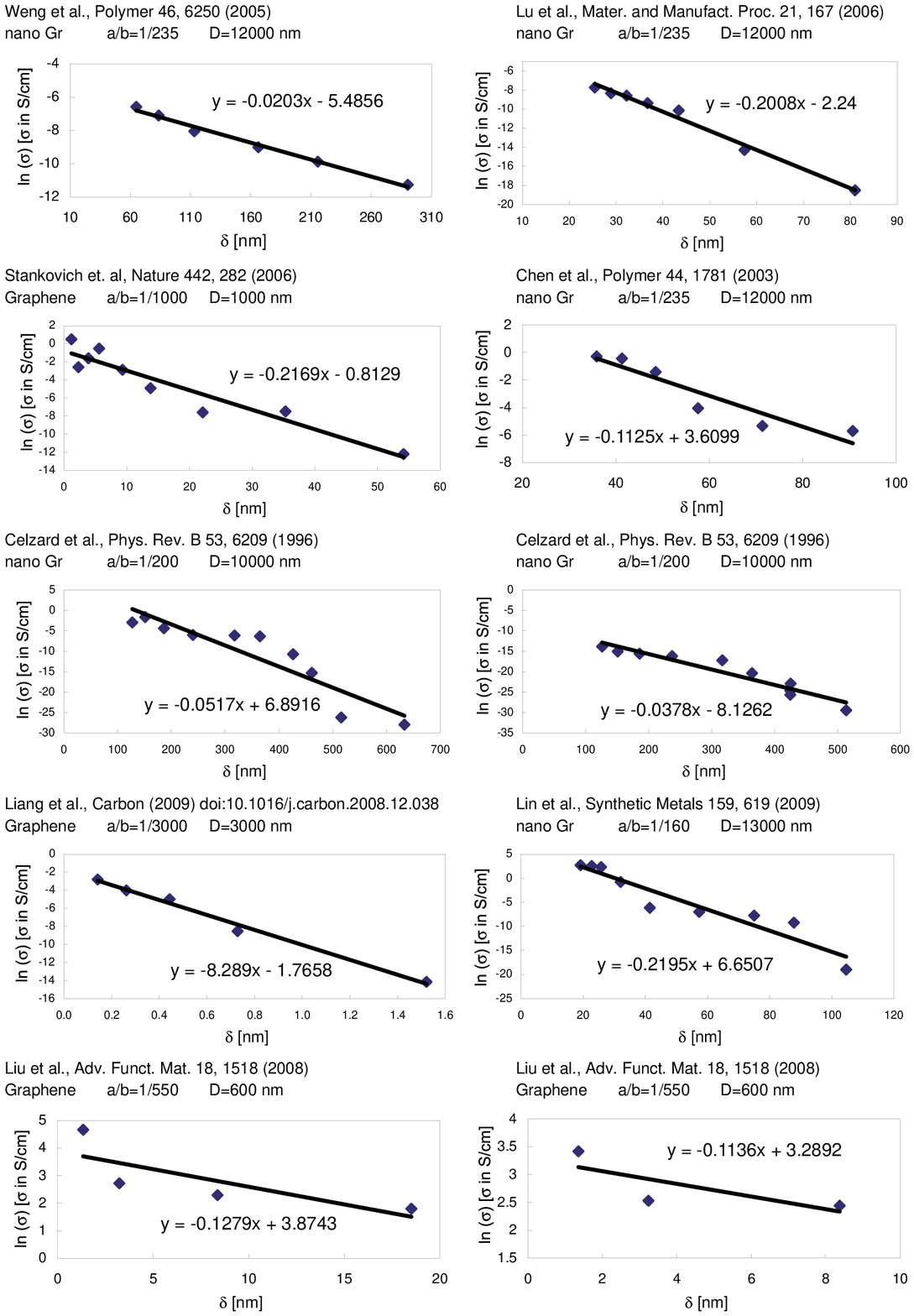}
  \caption{plots of $\ln\sigma$ as a function of $\delta$ for different polymer-nanosheet composites }\label{fig:SMfig11}
  \end{center}
\end{figure}

\begin{figure}[h!]
    \begin{center}
  \includegraphics[scale=1.0, clip='true']{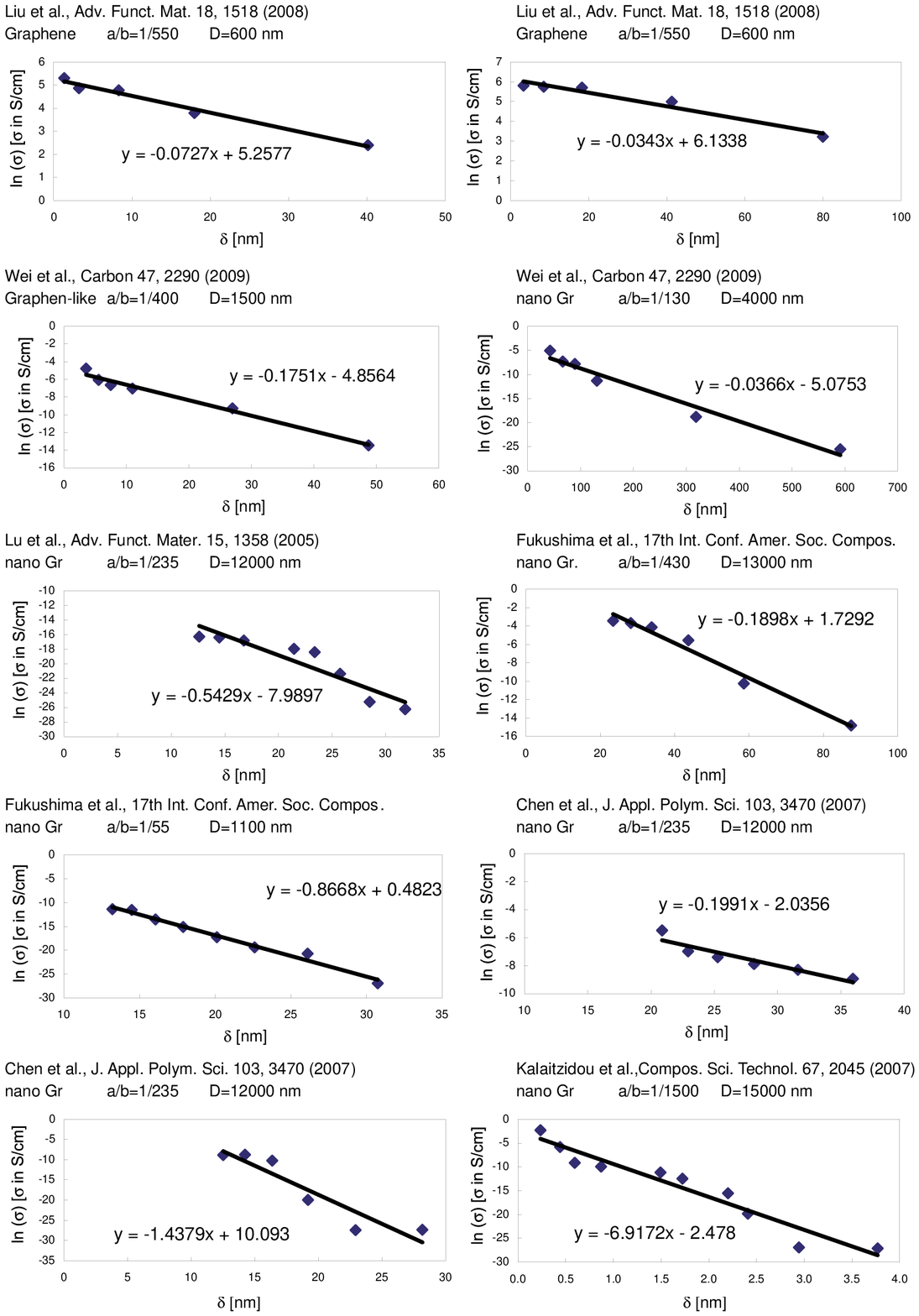}
  \caption{plots of $\ln\sigma$ as a function of $\delta$ for different polymer-nanosheet composites (cont.)}\label{fig:SMfig12}
  \end{center}
\end{figure}


\begin{thebibliography}{38}
\bibitem{notecross}
For the case in which there exist some electronic contribution to the
matrix conductivity, the total conducivity $\sigma_{\rm tot}$ is only
approximately the sum of  $\sigma$ and $\sigma_{\rm m}$. In this case
a more precise esimate of $\sigma_{\rm tot}$ could be obtained for
example from an effective medium approximation.
\end{thebibliography}
\end{document}